\documentclass[%
 aip,
 amsmath,amssymb,
 reprint,%
]{revtex4-1}
\usepackage{graphicx}
\usepackage{subfigure}
\usepackage{dcolumn}
\usepackage{bm}
\usepackage{color}
\usepackage[T1]{fontenc}
\usepackage{mathptmx}
\usepackage{multirow}
\usepackage{array}
\usepackage{float}
\usepackage{url}
\usepackage{CJKutf8}
\usepackage{pifont}
\usepackage{epstopdf, epsfig}
\usepackage{titlesec}
\titleclass{\subsubsubsection}{straight}[\subsection]
\newcounter{subsubsubsection}[subsubsection]
\renewcommand\thesubsubsubsection{\thesubsubsection.\alph{subsubsubsection}}
\titleformat{\subsubsubsection}
 {\normalfont\normalsize\bfseries}{\thesubsubsubsection}{1em}{}
\titlespacing*{\subsubsubsection}
{0pt}{3.25ex plus 1ex minus .2ex}{1.5ex plus .2ex}

\begin{document}
\preprint{AIP/123-QED}

\title{Droplet coalescence kinetics: thermodynamic non-equilibrium effects and entropy production mechanism}

\author{Guanglan Sun \begin{CJK*}{UTF8}{gbsn} (孙光兰) \end{CJK*}}
 \affiliation{School of Liberal Arts and Sciences,
Hebei Key Laboratory of Trans-Media Aerial Underwater Vehicle,
North China Institute of Aerospace Engineering, Langfang 065000, China}
\affiliation{School of Physics, Beijing Institute of Technology, Beijing 100081, China}
\author{Yanbiao Gan \begin{CJK*}{UTF8}{gbsn} (甘延标) \end{CJK*}}
 \thanks{Corresponding author: gan@nciae.edu.cn}
  \affiliation{School of Liberal Arts and Sciences,
Hebei Key Laboratory of Trans-Media Aerial Underwater Vehicle,
North China Institute of Aerospace Engineering, Langfang 065000, China}
\author{Aiguo Xu \begin{CJK*}{UTF8}{gbsn} (许爱国) \end{CJK*}}
 \thanks{Corresponding author: Xu\_Aiguo@iapcm.ac.cn}
\affiliation{Laboratory of Computational Physics, Institute of Applied Physics and Computational Mathematics, P. O. Box 8009-26, Beijing 100088, P.R.China}
\affiliation{HEDPS, Center for Applied Physics and Technology, and College of Engineering, Peking University, Beijing 100871, China}
\affiliation{State Key Laboratory of Explosion Science and Technology, Beijing Institute of Technology, Beijing 100081, China}
\affiliation{National Key Laboratory of Shock Wave and Detonation Physics, Mianyang 621999, China}
\author{Qingfan Shi \begin{CJK*}{UTF8}{gbsn} (史庆藩)\end{CJK*}}
\affiliation{School of Physics, Beijing Institute of Technology, Beijing 100081, China}

\date{\today}

\begin{abstract}
The thermodynamic non-equilibrium (TNE) effects and the relationships between various TNE effects and entropy production rate, morphology, kinematics, and dynamics during two initially static droplet coalescence are studied in detail via the discrete Boltzmann method.
The temporal evolutions of the total TNE strength $\bar D^*$ and the total entropy production rate can both provide concise, effective and consistent physical criteria to distinguish the stages of droplet coalescence. Specifically, when the total TNE strength $\bar D^*$ and the total entropy production rate reach their maxima, it corresponds to the time when the liquid-vapor interface length changes the fastest;
when the total TNE strength $\bar D^*$ and the total entropy production rate reach their valleys, it corresponds to the moment of the droplet being the longest elliptical shape.
During the merging process, the force contributed by surface tension in the coalescence direction acts as the primary promoting force for droplet coalescence and reaches its maximum concurrently with coalescent acceleration. In contrast, the force contributed by non-organized momentum fluxes (NOMFs) in the coalescing direction inhibits the merging process and reaches its maximum at the same time as the total TNE strength $\bar D^*$.
For the coalescence of two unequal size droplets, contrary to the larger droplet, the smaller droplet exhibits larger values for total TNE strength $\bar D^*$, merging velocity, driving force contributed by surface tension, and resistance contributed by the NOMFs. Moreover, these values gradually increase with the initial radius ratio of the large and small droplets due to the stronger non-equilibrium driving forces generated by larger curvature. However,  non-equilibrium components and forces related to shear velocity in the small droplet, are all smaller than those in the larger droplet and gradually decrease with the radius ratio.
This study offers kinetic insights into the complexity of thermodynamic non-equilibrium effects during the process of droplet coalescence, enhancing our understanding of the underlying physical processes in both engineering applications and the natural world.
\end{abstract}

\maketitle

\section{\label{sec:level1} Introduction}
Droplet coalescence and collision are fundamental phenomena with widespread applications in diverse fields such as nature, agricultural production, engine combustion systems, bioscience, and microfluidics. In nature, the collision, bouncing, and merging of raindrops are core mechanisms of raindrop growth and thunderstorm development. These processes have been a subject of continuous attention in atmospheric science since the early research of Benjamin Franklin and Lord Raleigh\cite{RN176,RN177}. In industrial settings, the coalescence of droplets is the main process that determines the appearance and shelf life of emulsifier products, like salad dressing and mayonnaise\cite{RN178}. In dense spray systems, such as those used in combustion applications, the strong interaction of droplets, including collision, rupture and coalescence significantly impact combustion processes and flame  temperature\cite{RN179,RNli2021,RN182}. In the field of life sciences, researchers have also identified striking similarities between the interaction between membrane, vesicle, and nucleolar fusion and droplet fusion\cite{RN123,RN184}. Controllable phase fusion has received widespread attention in the field of microfluidics\cite{ChenR2021POF,RN160,LiuHh2011,RN187,RN188,RN189,RN190,RN191}. There are four main methods for the coalescence of droplets in microfluidic devices: using channel geometry,  taking advantage of specific properties of liquids, making the use of thermal capillary effect, and utilizing microelectrodes to control droplet coalescence\cite{RN187}. For instance, T-shaped structures and other entangled structures are employed to induce the fusion of two droplets in microfluidic channels\cite{RN189,RN190}.
Overall, droplet coalescence and collision play a crucial role in diverse fields, and studying the merging mechanisms holds significant scientific and practical value.

The coalescence of droplet is a complex phenomenon that involves physical mechanisms at various scales, and it has been extensively studied through experiments\cite{RN127,RN185,RN186,RN187,RN188,RN189,RN190,RN191,RN192,RN193,RN194,RN195,RN196,RN197,RN198,RN199,
RN200,RN201,RN202,RN203,RN204,RN205,RN206,RN207,RN208,RN209,RN228,RN229,RN230}, theoretical analyses\cite{RN210,RN211,RN212,Succi2023,FeiLL2021POF,LiangH2016,RN216,RN217,RN218}, and numerical simulations\cite{RN9,Karlin2015,RN75,RN219,RN220,Yeomans2007,QinFF2018POF,RN222,YangZR2022,RN223,RN224,RN225,RN226,RN227,WenBH2023,HOSSEINI20231}. Droplet merging can generally be classified into two types: dynamic merging and static merging.  The study of droplet coalescence focuses on three main parameters: Bond number  ${\rm{Bo}} = {{\Delta \rho gR} \mathord{\left/
 {\vphantom {{\Delta \rho gR} \sigma }} \right.
 \kern-\nulldelimiterspace} \sigma }$  ,  Weber number  ${\rm{We}} = {{\rho {v^2}R} \mathord{\left/
 {\vphantom {{\rho {u^2}R} \sigma }} \right.
 \kern-\nulldelimiterspace} \sigma }$ and  Ohnesorge number ${\rm{Oh}} = {\mu  \mathord{\left/
 {\vphantom {\mu  {\sqrt {\rho R\sigma } }}} \right.
 \kern-\nulldelimiterspace} {\sqrt {\rho R\sigma } }}$. Here, $\rho $ represents the density of liquid phase, $R$ is the radius of the droplet, $u$ is the droplet velocity, $\Delta \rho$ is the density difference between the gas and liquid phases, $\sigma$ is surface tension at the interface, $\mu$ is the coefficient of viscosity, and $g$ is the acceleration of gravity. Understanding these parameters is of crucial scientific significance and practical importance in comprehending the dynamics of droplet coalescence across a wide range of applications.

Researches on dynamic droplet merging primarily focus on understanding the collision and fusion dynamics of droplets. The process of collision-coalescence behavior between two droplets can be roughly divided into the following distinct stages\cite{RN228,RN229,RN230}: (i) Approach: initially, two liquid droplets draw closer to each other; (ii) Contact and Deformation: as they approach, the droplets make contact and undergo deformation; (iii) Interfacial Liquid Film Discharge: the interfacial liquid film between the droplets undergoes a discharge process; (iv) Post-Collision Process: subsequently, the two droplets may either coalesce, adhere without coalescence, or rebound. The post-collision merging process can be further dissected into three stages: (a) Boundary Layer Interface Rupture and Reconnection Phase: this phase involves the rupture and reconnection of the boundary layer interface; (b) Convergence and Integration Stage: the droplets enter a stage of convergence and integration; (c) Damped Oscillation of Merged Droplet: following the merger, a new droplet undergoes a damped oscillation process. In 1977, Qian and Law\cite{RN203} conducted an extensive experimental study on the collision behavior of two droplets at various Weber numbers and density ratios. Based on collision parameters and the Weber number, they identified five collision outcomes:
(i) the two droplets collide and merge after small deformation; (ii) the two droplets bounce back after collision; (iii) the two droplets merge after collision and deformation; (iv) the two droplets are separated after merging in near positive collision; and (v) the two droplets are separated after eccentric collision and merger.

The focuses of the research of initial static droplets coalescence are the process of liquid membrane drainage during droplet coalescence, the growth process of the radius of the bridge with negative curvature, and the damped oscillation process after the merging of two droplets. By studying the coalescing process of initial static droplets, the basic process and principle of thin film drainage and construction and change of liquid bridge can be observed in detail, which provides a foundation for active controlling fields such as droplet micromixing, microreaction, and microfluidics. Upon the merging of two droplets, a liquid bridge forms at their initial contact point, creating a meniscus structure along the outer boundary of the bridge. The strong negative curvature near the point of contact, as dictated by Young-Laplace's law, results in a notable pressure gradient within the upper region of the meniscus. This gradient, driven by differences in surface tension at the interface, prompts a convergence toward the intermediate contact point, leading to an increase in the bridge's radius.
Hopper\cite{RN210,RN217}conducted pioneering theoretical work on the evolving two-dimensional topology of merging droplets. He gave an exact analytical solution for the coalescence of two infinitely long two-dimensional viscous liquid cylinders (two circular droplets) in the Stokes limit. Eggers \emph{et al.} \cite{RN211} demonstrated the applicability of Hopper's solution in the case of three-dimensional spherical droplets. This high-viscosity mechanism is known as the Stokes regime, where viscous force dominates the coalescence process.
The evolution relationship of liquid bridge radius ${r_b}$ over time is ${r_b} \sim {C_{\rm{V}}}t\ln t$ \cite{RN200,RN201,RN202,RN211,RN218,RN222}. When the viscosity of the system is low enough, the surface tension inertia dominates the fusion process. This mechanism is called the inviscid regime. Under the inertia mechanism, the evolution relationship of the liquid bridge radius ${r_b}$ over time is ${r_b} \sim {C_{\rm{I}}}{t^{{1 \mathord{\left/ {\vphantom {1 2}} \right.
 \kern-\nulldelimiterspace} 2}}}$\cite{RN127,RN196,RN200,RN201,RN202,RN211,RN222,RN231}. Paulsen \emph{et al.} \cite{RN201,RN202} believed that the initial state of droplet merging must be a state of equilibrium between viscous force and capillary inertia force, in which neither viscous force nor inertial force can be ignored. This mechanism is called inertia limited viscous mechanism (ILV), under which ${r_b}$ evolves over time as ${r_b} \sim {C_{{\rm{ILV}}}}t$. Here, ${C_ {\rm {V}}} $, ${C_ {\rm {I}}} $and ${C_ {{\rm {ILV}}}} $ are the growth factor of ${r_b}$ under three kinds of mechanism ,respectively. Although they have different expressions, they are positively correlated with the surface tension $\sigma $ and negatively correlated with the viscosity coefficient $\mu $. In summary, for highly viscous systems dominated by viscosity (${\rm{Oh}} \ge {\rm{1}}$), the coalescence process commences with the inertial finite viscous mechanism and dynamically transitions to the Stokes mechanism. In contrast, for systems with lower viscosity and dominated by inertia (${\rm{Oh}} < {\rm{1}}$), the coalescence begins with the inertially finite viscosity mechanism and then transitions dynamically to the inertially finite viscosity mechanism.
\begin{figure*}[htbp]
{\centering
\includegraphics[width=1.\textwidth]{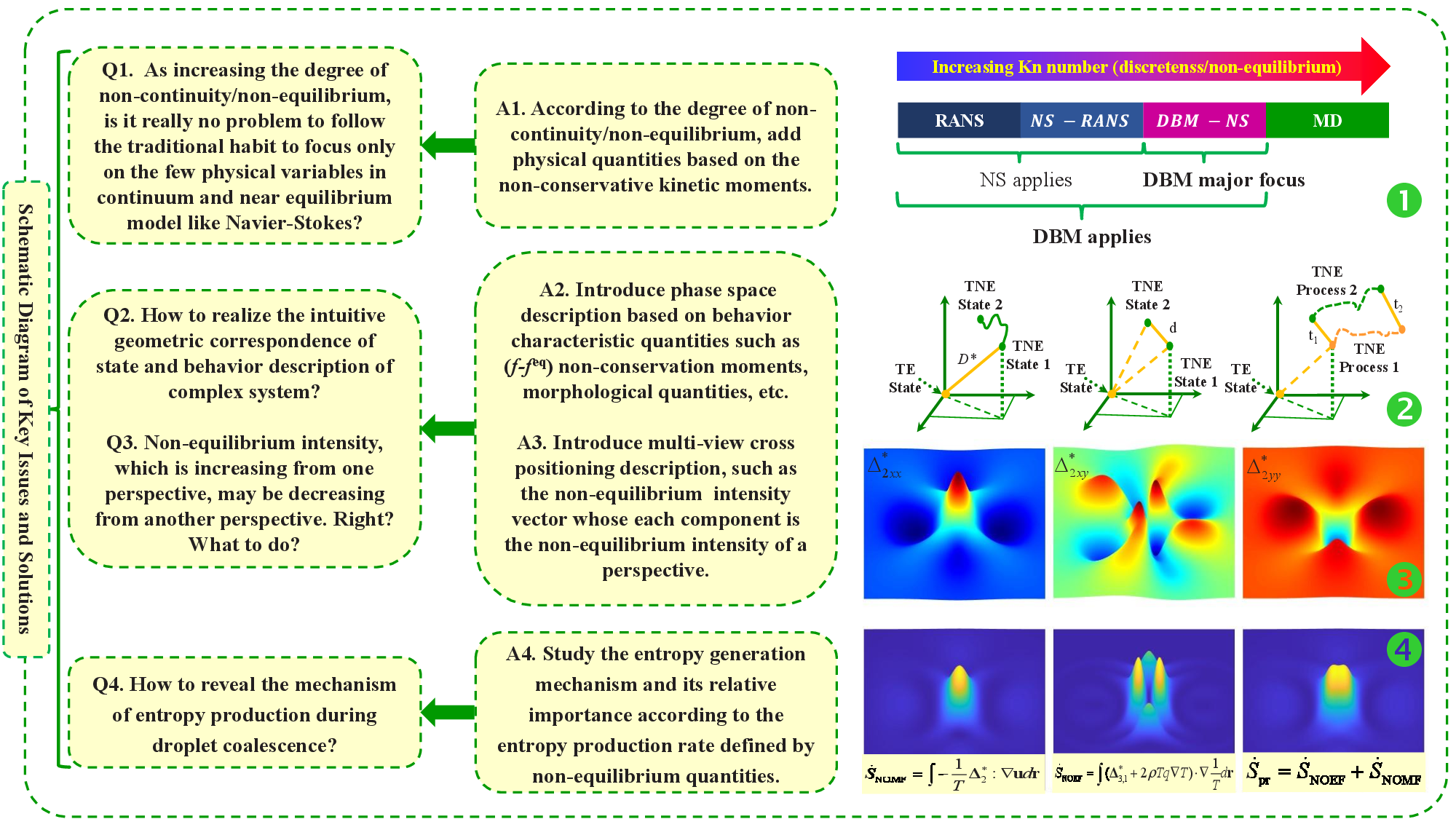}}
 \caption{\centering{Schematic diagram of key issues and solutions in non-equilibrium flow research. In this figure, DBM is an abbreviation for the recently developed discrete Boltzmann \emph{modeling and analysis} method;
   Line \ding{183} contains a schematic representation of the non-equilibrium phase space constructed based on the non-conservative moments $(f-f^{eq})$, providing an intuitive geometric correspondence for the study of non-equilibrium states and effects;
   Line \ding{184} depicts the complex spatial distributions of the components of the second-order non-equilibrium measures at a certain moment during the droplet merging process (the quadrupole and octupole structures);
    Line \ding{185} illustrates the spatial distributions of entropy production rates associated with two distinct non-equilibrium effects and the total entropy production rate at a specific moment during the droplet coalescence\cite{Xu2023brief}.
   }}
 \label{FSD}
\end{figure*}

Despite significant progress has been made in the study of droplet merging process, the current numerical study of this topic faces three prominent challenges:
(i) \emph{Kinetic modeling and analysis of non-equilibrium effects in droplet merging processes.} Previous numerical simulations of droplet coalescence were primarily based on macroscopic methods, say the Navier-Stokes equations. Macroscopic methods are effective in capturing the large-scale behaviors during the merging process. However, they are insufficient to provide intricate descriptions of small-scale structures, such as liquid bridges, droplet boundary layers, vortices, droplet fragments or splashes, and surface fluctuations. In the vicinity of these small-scale structures, the system deviates significantly from thermodynamic equilibrium due to scale effects, resulting in high Knudsen numbers. The spatial gradients of macroscopic physical quantities (such as pressure, density, velocity, and temperature) and intermolecular interaction forces closely related to these phenomenons will cause complex thermodynamic non-equilibrium (TNE) effects. Additionally, in cases of rapid collision and merging process, the emergence of fast modes prevents the system returning to thermodynamic equilibrium timely\cite{YangY2022,Ezzatneshan2020PRE,RN90,RN203,Karlin2015}. These spatial and temporal non-equilibrium effects significantly challenge the effectiveness of macroscopic models. Therefore, it is imperative to develop multi-scale kinetic models to reveal the thermodynamic non-equilibrium effects during the droplet merging process (see line \ding{182} in Fig. \ref{FSD}).
(ii) \emph{Data analysis and information extraction scheme for complex physical fields during the droplet merging process.} The process of droplet coalescence, especially under the influence of complex physical fields, involves rich and complex multi-scale spatiotemporal structures (see line \ding{184} in Fig. \ref{FSD}). How to effectively extract feature structures at different levels from massive simulation data and reveal the impact of feature structures on heat transfer and flow behavior remains an open topic.
(iii) \emph{The mechanism of entropy production during droplet merging remains unclear.}
Entropy production, a highly concerned physical quantity in the scientific and engineering realms, reflects the thermal equilibrium tendency, irreversibility, heat transfer direction, and internal disorder of a system during its evolution. A thorough understanding of entropy production in the droplet merging process is crucial for enriching the thermodynamic theory of droplet coalescence, ultimately aiding in optimizing and controlling the merging phenomenon. This is particularly pertinent in engineering applications such as droplet spraying, coating, combustion, etc. However, the non-equilibrium entropy production mechanism within the droplet coalescence process is currently not well-defined. Figure \ref{FSD} displays the schematic diagrams for the three key issues and their solutions.

Figure \ref{FSD} manifests that the recently developed discrete Boltzmann method (DBM) serves as \emph{a physical modeling and the complex physical field analysis method}. In addition to the description function equivalent to the corresponding level of the extended hydrodynamic equations, DBM also provides a set of complex physical field analysis methods, particularly in investigating TNE behaviors\cite{Xu2022CMK,Zhang2023PoF,Zhang2023CAF,Zhang2022PoF,RN22,RN29,RN30,RN32,RN78,Lai2023,Lai2016PRE,LinCD2023,LinCD2017,RN81,RN82,RN84,
RN86,RN90,RN92,RN25,Chen2022PRE,ChenF2018,ChenF2020,ZhangYD2023,QiuRF2021PRE,QiuRF2022POF,RN107}. In other words, in terms of numerical simulation research, DBM is not only responsible for pre-simulation, but also for post-simulation.
A pivotal point in DBM research began with Xu  \emph{et al.}'s work in 2012\cite{RN22}, which proposed using non-conservative moments of $(f - {f^{eq}})$ to quantify system deviations from the thermodynamic equilibrium (and/or continuum case) and to detect the resulting various kinds of effects. Each individual component of the non-conserved moment of $(f - {f^{eq}})$  describes the specific deviation (from equilibrium/continuum)  and the resulting non-equilibrium/discrete effects from its own perspective. It is clear that each of these perspective contributes a non-equilibrium/discrete strength/degree/intensity. The non-equilibrium/discrete strengths from different perspectives are interrelated and complementary to each other, but they cannot replace each other. For complex flows, the non-equilibrium/discrete strength may increase from one perspective but decreases from the other perspective. The one single perspective description of non-equilibrium/discrete strength may introduce confusion. Therefore, the non-equilibrium/discrete strength vector is introduce the multiplex locate the non-equilibrium/discrete behavior\cite{Xu2022CMK,Zhang2022PoF}. Since some higher order moments of $(f - {f^{eq}})$ do not correspond in traditional fluid theory, people may feel them unfamilar. To solve this problem, Xu  \emph{et al.} proposed the phase space description method based on the non-conserved kinetic moments of $(f - {f^{eq}})$ to give an intuitive geometric correspondence to the non-equilibrium/discrete state and effects \cite{Xu2022CMK,RN32,RN92,RN81}. Soon later, the phase space description method was extended to describe any set of characteristic behaviors to be studied\cite{Xu2022CMK,RN81}.
A point within the phase space corresponds to a specific set of characteristic behaviors of the system. Distance concepts in the phase space or its sub-spaces are used to describe the difference and similarity of behavior (see line \ding{183} in Fig. \ref{FSD}).
For example, Gan \emph{et al.} \cite{RN29} used a DBM of multiphase flow to study the thermohydrodynamic non-equilibrium (THNE) and TNE effects in the phase separation process. They defined TNE strength and discovered that the time evolution of the TNE intensity provides a convenient and efficient physical criterion for distinguishing between the stages of spinodal decomposition and domain growth. Zhang \emph{et al.}\cite{RN90} studied the coalescence and collision of two droplets under isothermal conditions by using DBM of multiphase flow based on Enskog collision model.
They found that during the coalescence process, the strength of the second order non-equilibrium central moment, $\bar D_2^*$, is always significantly larger than the strength of the third order non-equilibrium central moment, $\bar D_3^*$. Moreover, it was observed that $\bar D_2^*$ can be used to identify the different stages of the collision process and to distinguish different types of collisions. In the study of TNE effects during bubble coalescence\cite{RN25}, it is found that the two characteristic moments provided by the time evolution of the unorganized momentum flow in the merging direction can be used to divide the bubble coalescence. More importantly, the total non-equilibrium intensity, coalescing acceleration and the rate of reduction in liquid-vapor boundary length are highly correlated with their magnitudes reaching their maximum values almost simultaneously.

In this paper, the TNE behaviors, the influences of TNE and the entropy production rate during the merging process of two initial static droplets are investigated by a thermal multiphase DBM.
Section II introduces the underlying physical model.
Section III encompasses simulations and analyses, covering non-equilibrium characteristics and their influences on droplet coalescence, the effects of the initial radius ratio of the larger and smaller droplets, as well as the entropy production rate during the droplet coalescence. Section IV draws the conclusions derived from our findings.
\section{Multiphase DBM  with TNE effects}
The DBM serves as a methodology for constructing physical models and analyzing complex physical fields based on discrete Boltzmann equation. DBM primarily targets the investigation of mesoscale behavior, an area that challenges macroscopic continuous modeling due to infeasibility or inconvenience, and is similarly out of reach for microscopic molecular dynamics simulations due to scale limitations.
 As a \emph{modeling and analysis} method, DBM mainly has two capabilities: physical modeling and complex physics field analysis.

\emph{How to conduct physical modeling}? The initial step is the modification and simplification of the Boltzmann equation. The evolution equation of molecular velocity distribution function is
\begin{equation}
{\frac{{\partial f}}{{\partial t}}}+\mathbf{v}\cdot {\frac{{\partial f}}{{%
\partial \mathbf{r}}}}+\mathbf{a}\cdot {\frac{{\partial f}}{{\partial
\mathbf{v}}}}={({\frac{{\partial f}}{{\partial t}}})_{c}},
\end{equation}
where $f=f({\mathbf{r},\mathbf{v},t})$ is the molecular
velocity distribution function, $\mathbf{r}$, $\mathbf{v}$, and $\mathbf{a}$ are the position space coordinate, the velocity space coordinate, and the acceleration generated by the total extra force, respectively. $({\partial f}/{\partial t})_c$ represents the collision term. To linearize the collision term as $({\partial f}/{\partial t})_c = -{({f-{f^{eq}}})}/{\tau}$ \cite{BGK1954}, the local equilibrium state function ${f^{eq}}$ is introduced on the basis of satisfying the constraint conditions of $\int {\bm{\Psi }{({\partial f}/{\partial t})_c}}d\mathbf{v} = \int -\bm{\Psi }{({f-{f^{eq}}})}/{\tau} d\mathbf{v}$. Here $\bm{\Psi}=[1,\mathbf{v},\mathbf{vv}, \mathbf{vvv}]^{Tr}$, $\tau$ is a relaxation time and ${f^{eq}}=\rho {({1/2\pi T})}\exp[-{{(\mathbf{v}-\mathbf{u})}^{2}}/2T] $ represents the two-dimensional Maxwellian distribution function on the Bhatnagar-Gross-Krook (BGK) model.

To describe non-ideal fluids and surface tension effects, an appropriate force term\cite{RN112} $I=-[A+{\bf B}\cdot ({\bf v}-{\bf u})+(C+{C_{q}}){({\bf v}-{\bf u})^{2}}]{f^{eq}}$ is introduced in the right hand of the Boltzmann-BGK equation,
\begin{equation}
{\frac{{\partial f}}{{\partial t}}}+\mathbf{v}\cdot {\frac{{\partial f}}{{%
\partial \mathbf{r}}}}=-{\frac{1}{\tau }}(f-{f^{eq}})+I.
\end{equation}
Here, $\rho $, $\mathbf{u}$, and $T$ represent the local density, the velocity, and the temperature, respectively. $A$ is defined as $A=-2\left( {C+{C_{q}}}\right)T$ to guarantee mass conservation; $\mathbf{B}$ is given by $\mathbf{B}={\frac{1}{{\rho T}}}\bm {\nabla} \cdot \left[ {\left( {{P^{\mathrm{CS}}-}\rho T}\right) \mathbf{I}+\bm{\Lambda }}\right]$, its first term accounts for deviations from the non-ideal fluid equation of state (EOS) and the second term considers the contributions of the density and temperature gradients to pressure tensor with $\bm{\Lambda }=K\bm {\nabla} \rho
\bm {\nabla} \rho -K(\rho {\nabla ^{2}}\rho +{{{{\left\vert {\bm {\nabla} \rho }%
\right\vert }^{2}}}/2})\mathbf{I}-[\rho T\bm {\nabla} \rho \cdot \bm {\nabla} (K/T)]\mathbf{I}$,  $\mathbf{I}$ being a unit tensor, and $K$ being the surface tension coefficient; $C = \frac{1}{{2\rho {T^2}}}\{ \left( {{P^{{\rm{CS}}}} - \rho T} \right)\bm {\nabla}  \cdot {\bf{u}} + {\bm{\Lambda }}:\bm {\nabla} {\bf{u}} + a{\rho ^2}\bm {\nabla}  \cdot {\bf{u}} - K\left[ {\frac{1}{2}\bm {\nabla} \rho  \cdot \bm {\nabla} \rho \bm {\nabla}  \cdot {\bf{u}} + \rho \bm {\nabla} \rho  \cdot \bm {\nabla} \left( {\bm {\nabla}  \cdot {\bf{u}}} \right) + \bm {\nabla} \rho  \cdot \bm {\nabla} {\bf{u}} \cdot \bm {\nabla} \rho } \right]\} $ represents a partial contribution to energy;  and ${C_{q}}={\frac{1}{{\rho {T^{2}}}}}\bm {\nabla} \cdot \left( {q\rho T\bm {\nabla} T}\right)$  accounts for the heat conduction within the system. The Prandtl number $\Pr ={c_{p}\mu/{\kappa_T}}={\tau/{ {(\tau  - q)}}}$ can be adjusted by modulating the parameter $q$ in ${C_{q}}$. Here $\mu=\rho T\tau$, ${\kappa_T}=c_{p}\rho T(\tau  - q)$ and $c_{p}$ are the viscosity coefficient, the heat conductivity and the isobaric heat capacity, respectively. $P^{\mathrm{CS}}$ stands for the Carnahan--Starling EOS ${P^{\mathrm{CS}}}=\rho T{\frac{{1+\eta +{\eta ^{2}}-{\eta ^{3}}}}{{{{(1-\eta)}^{3}}}}}-a{\rho ^{2}}$ with $\eta ={{b\rho }/4}$, and $a$ and $b$ being the attraction and repulsion parameters, respectively.

The definition of any non-equilibrium strength depends on the research perspective.
Here, we choose the kinetic moments $\bm{\Theta }_{n} = (\mathbf{M}_0, \mathbf{M}_1, \mathbf{M}_{2,0}, \mathbf{M}_2, \mathbf{M}_{3}, \mathbf{M}_{3,1}, \mathbf{M}_{4,2})^{Tr}$ that the discrete equilibrium distribution function should satisfy during the coarse-grained physical modeling process, where $\mathbf{M}_{m,n}^{eq}=\int_{-\infty }^{\infty } (\frac{1}{2})^{1-\delta
_{m,n}}f^{eq}{\underbrace{{\mathbf{vv}}\cdots {\mathbf{v}}}_{n}(\mathbf{v\cdot v})}^{\frac{m-n}{2}}d\mathbf{v}$, $\delta_{m,n}$ is the Kronecker delta function.
The kinetic moment set provides a distinctive research perspective to explore the non-equilibrium behaviors of interest.
The modeling accuracy and research perspective should be adjusted and improved in a timely manner\cite{RN107}.
These moments actually are physical constraints for the discretization of velocity space, $\int {{f^{eq}}\bm{\Psi }\left(\mathbf{v}\right) }d\mathbf{v}=\sum_{l}{{f_{l}}^{eq}\bm{\Psi }%
\left( {{\mathbf{v}_{l}}}\right) }$.
After discretization of the velocity space, the modified discrete Boltzmann-BGK equation takes the following form
\begin{equation}
{\frac{{\partial {f_{ki}}}}{{\partial t}}}+{\mathbf{v}_{ki}}.{\frac{\partial
}{{\partial \mathbf{r}}}}{f_{ki}}=-{\frac{1}{\tau }}({f_{ki}}
-f_{ki}^{eq})+{I_{ki}},
\end{equation}
where $f_{ki}^{eq}$ (see the Appendix) represents the discrete counterpart of the local equilibrium
distribution function, and ${I_{ki}}=-[A+{\mathbf B}\cdot ({\mathbf v}_{ki}-{\mathbf u })+(C+{C_{q}}){({
\mathbf v_{ki}}-{\mathbf u})^{2}}]f_{ki}^{eq}$ .

Taking the moments of equation (3) with the collision invariant vector $1$, ${\mathbf{v}_{ki}}$, and ${{{v}_{ki}^{2}}/2}$, the generalized thermohydrodynamic equations for the CS non-ideal fluids with the surface tension effect are obtained \cite{RN40}
\begin{equation}
{\frac{{\partial \rho }}{{\partial t}}}+\bm {\nabla} \cdot (\rho \mathbf{u})=0,
\end{equation}
\begin{equation}
{\frac{{\partial (\rho \mathbf{u})}}{{\partial t}}}+\bm {\nabla} \cdot (\rho
\mathbf{uu}+P^{\rm{CS}}\mathbf{I})+\bm {\nabla} \cdot (\bm{\Lambda }+\bm{\Delta }%
_{2}^{\ast })=0,
\end{equation}
\begin{eqnarray}
\begin{aligned}
{\frac{{\partial {e_{T}}}}{{\partial t}}}+ \bm {\nabla} \cdot ({e_{T}}\mathbf{u}+P^{\rm{CS}}\mathbf{u})\qquad \qquad  \qquad \quad \quad \ \quad \, \, \, \qquad \\
+\bm {\nabla} \cdot \left[ {(\bm{\Lambda }+\bm{\Delta }%
_{2}^{\ast })\cdot \mathbf{u}+\bm{\Delta }_{3,1}^{\ast }+2\rho Tq \bm {\nabla} T}\right] =0,
\end{aligned}
\end{eqnarray}
with ${e_{T}}=\rho T-a{\rho ^{2}}+{{K{{\left\vert {\bm {\nabla} \rho }\right\vert }^{2}}}/2+\rho u^{2}/2}$  the total energy density.
It should be emphasized that the ability to recover the hydrodynamic equations (HEs) mentioned above is only a part of the physical capabilities of the DBM. In fact, at a macroscopic level, DBM corresponds to the evolution equations for moments of the distribution function at all orders, encompassing both the evolution equations for low-order conservative moments (i.e., HEs), and high-order non-conservative moments\cite{RN107}. These equations are collectively referred to as the extended hydrodynamic equations. The significance of the extended part (i.e., the evolution equations for non-conservative moments) rapidly increases with the level of discretization/non-equilibrium.

\emph{How to  detect, describe, present, and analyze non-equilibrium states and effects}? DBM is capable of capturing non-equilibrium effects that contain behaviors of molecular average behavior and thermal fluctuation (the THNE effects), as well as non-equilibrium effects that only include molecular thermal fluctuation behavior (the TNE effects)\cite{RN107}. The local TNE effects defined  by the thermodynamic non-equilibrium moments $\bm{\Delta }_{m,n}^{*}$, which are defined as follows
\begin{equation}
{\bm{\Delta}}_{m,n}^{\ast }={\mathbf{M}}_{m,n}^{\ast }-{\mathbf{M}}%
_{m,n}^{\ast eq}.
\end{equation}
Among which, the four typical TNE measures with clear physical interpretations are
\begin{equation}
{\bm{\Delta }}_{2}^{\ast }={\mathbf{M}}_{2}^{\ast }-{\mathbf{M}}_{2}^{\ast
eq}=\sum\limits_{ki}{({\mathbf{v}_{ki}}-\mathbf{u})({\mathbf{v}_{ki}}-%
\mathbf{u})({f_{ki}}-f_{ki}^{eq})},
\end{equation}
\begin{equation}
{\bm{\Delta }}_{3}^{\ast }={\mathbf{M}}_{3}^{\ast }-{\mathbf{M}}_{3}^{\ast
eq}=\sum\limits_{ki}{({\mathbf{v}_{ki}}-\mathbf{u})({\mathbf{v}_{ki}}-%
\mathbf{u})({\mathbf{v}_{ki}}-\mathbf{u})({f_{ki}}-f_{ki}^{eq})},
\end{equation}
\begin{equation}
{\bm{\Delta }}_{3,1}^{\ast }={\mathbf{M}}_{3,1}^{\ast }-{\mathbf{M}}%
_{3,1}^{\ast eq}={\frac{1}{2}}\sum\limits_{ki}{{{({\mathbf{v}_{ki}}-\mathbf{u%
})}^{2}}({\mathbf{v}_{ki}}-\mathbf{u})({f_{ki}}-f_{ki}^{eq})},
\end{equation}
\begin{eqnarray}
\begin{aligned}
{\bm{\Delta }}_{4,2}^{\ast }={\mathbf{M}}_{4,2}^{\ast }-{\mathbf{M}}%
_{4,2}^{\ast eq} \qquad \qquad \qquad \qquad \quad \quad \, \, \, \qquad \\
={\frac{1}{2}}\sum\limits_{ki}{{{({\mathbf{v}_{ki}}-\mathbf{u%
})}^{2}}({\mathbf{v}_{ki}}-\mathbf{u})({\mathbf{v}_{ki}}-\mathbf{u})({f_{ki}}%
-f_{ki}^{eq})}.
\end{aligned}
\end{eqnarray}
${\bm{\Delta }}_{2}^{\ast }$, ${\bm{\Delta }}_{3}^{\ast }$, ${\bm{\Delta }}_{3,1}^{\ast }$ and ${\bm{\Delta }}_{4,2}^{\ast }$ correspond to non-organized momentum fluxes (NOMFs)/viscous stress, stress fluxes, non-organized energy fluxes (NOEFs)/heat flux, and fluxes of heat flux, respectively. The first-order analytical solutions for $\bm{\Delta }_{2}^{\ast }$ and $\bm{\Delta }_{3,1}^{\ast }$  are $\bm{\Delta }_{2}^{\ast (1)}=-\rho T\tau \left[{\bm {\nabla} {\bf u}+{{(\bm {\nabla} {\bf u})}^{Tr}}-\mathbf{I}\bm {\nabla} \cdot \mathbf{u}}\right]$, $\bm{\Delta }_{3,1}^{\ast (1)}=-2\rho T\tau \bm{\nabla} T$, respectively. In equations (7)-(11),
\begin{eqnarray}
\begin{aligned}
\mathbf{M}_{m,n}^{\ast }\left( {{f_{ki}}}\right) =\qquad \qquad \qquad \qquad \qquad  \qquad \quad \quad \ \quad \, \, \, \qquad\\
\sum\limits_{ki}{{\frac{1}{2}}^{(1-\delta_{m,n})}}{{f_{ki}}%
\overbrace{\underbrace{({\mathbf{v}_{ki}}-\mathbf{u})({\mathbf{v}_{ki}}-%
\mathbf{u})\cdots ({\mathbf{v}_{ki}}-\mathbf{u})}_{n}{{\left\vert {({\mathbf{%
v}_{ki}}-\mathbf{u})}\right\vert }^{(m-n)}}}^{m}},
\end{aligned}
\end{eqnarray}
\begin{eqnarray}
\begin{aligned}
\mathbf{M}_{m,n}^{\ast eq}\left( {f_{ki}^{eq}}\right)=\qquad \qquad \qquad \qquad \qquad  \qquad \quad \quad \ \quad \, \, \, \qquad \\
\sum\limits_{ki}{%
{{\frac{1}{2}}^{(1-\delta_{m,n})}}f_{ki}^{eq}\overbrace{\underbrace{({\mathbf{v}_{ki}}-\mathbf{u})({\mathbf{v}%
_{ki}}-\mathbf{u})\cdots ({\mathbf{v}_{ki}}-\mathbf{u})}_{n}{{\left\vert {({%
\mathbf{v}_{ki}}-\mathbf{u})}\right\vert }^{(m-n)}}}^{m}},
\end{aligned}
\end{eqnarray}
where $m$ is the highest power of $(\mathbf{v}_{ki}-\mathbf{u})$ (thermo-fluctuations of molecules relative to $\mathbf{u}$) in the moment,
and $n$ is the tensor order after the non-equilibrium moment is contracted.
 When $m=n$, $\mathbf{M}_{m,n}^{\ast }=
\mathbf{M}_{m}^{\ast }$\cite{RN107}.
It is clear that each component of $\bm{\Delta}_{m,n}^{\ast }$ presents a TNE strength from its corresponding perspective.
Based on these TNE strengths, one can also compose TNE strength from a new angle.
To provide a more comprehensive representation of
non-equilibrium effects during the droplet coalescence process, it is more
appropriate to introduce a non-equilibrium strength vector, such as
\begin{equation}
\mathbf{S}_{\text{TNE}}=\{|\bm{\Delta}_{2}^{\ast}|,|\bm{\Delta}_{3,1}^{\ast }|,%
|\bm{\Delta}_{3}^{\ast }|,|\bm{\Delta}_{4,2}^{\ast }|,...,|\bm{\Delta}%
_{m,n}^{\ast }|,{D{^{\ast }}}\},
\end{equation}
which is composed of various independent
non-equilibrium quantities and their nonlinear combinations. Here, ${D{^{\ast
}=}}\sum_{m,n}|\bm{\Delta}_{m,n}^{\ast }|$ is the total TNE
intensity. This vector effectively cross-describes the ways, extents, and
processes of system deviations from equilibrium from multiple perspectives.

\section{ Numerical simulations and results }\label{Numerical simulations}
In the simulations, all parameters are non-dimensionalized by the actual physical quantities \cite{RN30}. The recovery of actual physical quantities from the numerical results is feasible when parameters in the EOS and the real fluid's surface tension coefficient are provided. The D2V33 DVM is adopted to discrete the phase space (see Section A in the Appendix), which has been validated by several benchmarks\cite{RN233}. The fast Fourier transform (FFT) scheme with $16$-th order is used to discretize the spatial derivatives, and the second-order Runge-Kutta finite difference scheme is utilized to solve the temporal derivative (see Sections B and C in the Appendix)\cite{RN83}.
It should be pointed out that the precision of the numerical format will 
affect the realization of the physical function of the model to some extent. The discretization scheme presented in this paper is the optimal choice for meeting the current research requirements, which can effectively ensure the numerical conservation of the total energy of the system and significantly refrain the spurious velocity around the liquid-vapor interface. Of course, the current discrete format is not meant to be a standardized or optimal template for all problems, such as the simulation of multiphase flow systems under shock waves.
The computational grids are set as ${N_x} \times {N_y} = 256 \times 256$ with the spatial step of $\mathrm{\Delta }x = \mathrm{\Delta }y = {1/128}$ and the time step is $\mathrm{\Delta }t = 0.0001$. The parameters \emph{a} and \emph{b} in the EOS are chosen as $a = 2.0$ and $b = 0.4$, which fix the critical point at ${T_c} = 1.88657$, ${\rho _c}= 1.3044$, and ${P_c} = 0.8832$.

The initial state of two stationary droplets positioned horizontally side by side is defined as
\begin{equation}
\begin{array}{l}
\rho (x,y)={\rho _l} -\frac{{({\rho _{l}}-{\rho _{g}})}}{2}\tanh \left[ {%
\frac{{\sqrt{{{(x-{x_{0\rm l}})}^{2}}+{{(y-{y_{0\rm l}})}^{2}}}-{r_{0\rm l}}}}{{0.5w}}}%
\right]  \\
\mathrm{\ \ \ \ \ \ \ \ \ \ \ \ \ \ \ }-\frac{{({\rho _{l}}-{\rho _{g}})}}{2}\tanh %
\left[ {\frac{{\sqrt{{{(x-{x_{0\rm r}})}^{2}}+{{(y-{y_{0\rm r}})}^{2}}}-{r_{0\rm r}}}}{{%
0.5w}}}\right] .
\end{array}
\label{eq30}
\end{equation}
Here, ${\rho _{l}}=2.0658$, ${\rho _{g}}=0.6894$ are the density of liquid
and vapor phases at a temperature of $T = 0.954{T_c} = 1.8$; $w$ is the width of the boundary layer; ${r_{0\rm l}}$ and ${r_{0\rm r}}$ are the initial radii of the left and right droplets, respectively. (${x_{0 \rm l}}$, ${y_{0\rm l}}$) denotes the center coordinate of the left droplet, and (${x_{0\rm r}}$, ${y_{0\rm r}}$) represents the center coordinate of the right droplet. The temperature of the system is allowed to evolve freely during the simulations. This initial state configuration provides a starting point for the study of droplet dynamics, interactions and TNE effects within the system.
\subsection{Non-equilibrium characteristics during the coalescence of two equal-size droplets}
\begin{figure*}[htbp]
{\centering\includegraphics[width=0.8\textwidth]{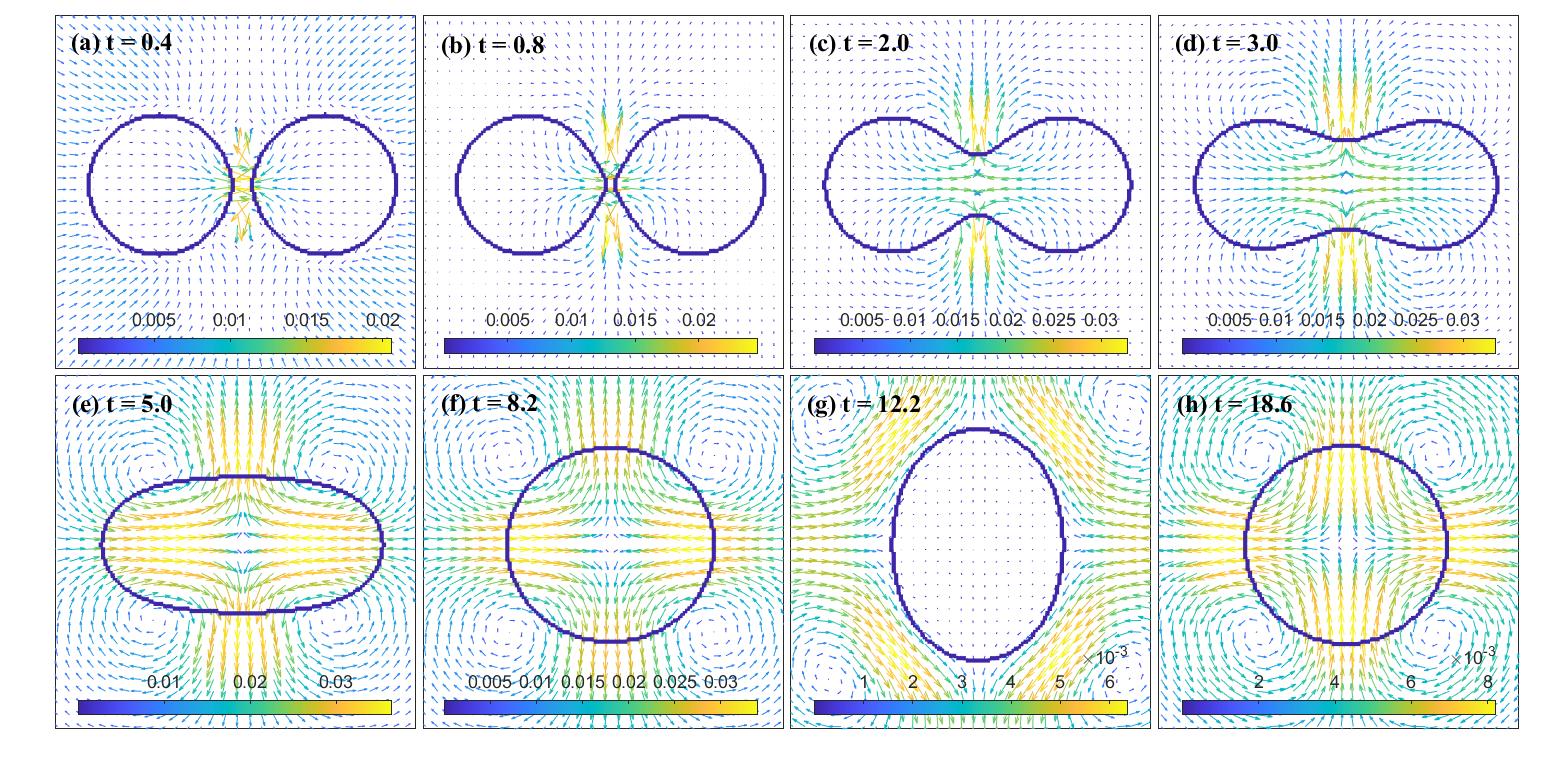}}
 \caption{\centering{Spatial distributions of velocity vectors at various time instances $t=0.4$, $0.8$, $2.0$, $3.0$, $5.0$, $8.2$, $12.2$ and $18.6$. Here, each sub-figure shows a partial simulated area with grid dimensions  ${N'_x} \times {N'_y} = 150 \times 150$.}}
 \label{F1}
\end{figure*}
\begin{figure*}[htbp]
{\centering
\includegraphics[width=0.8\textwidth]{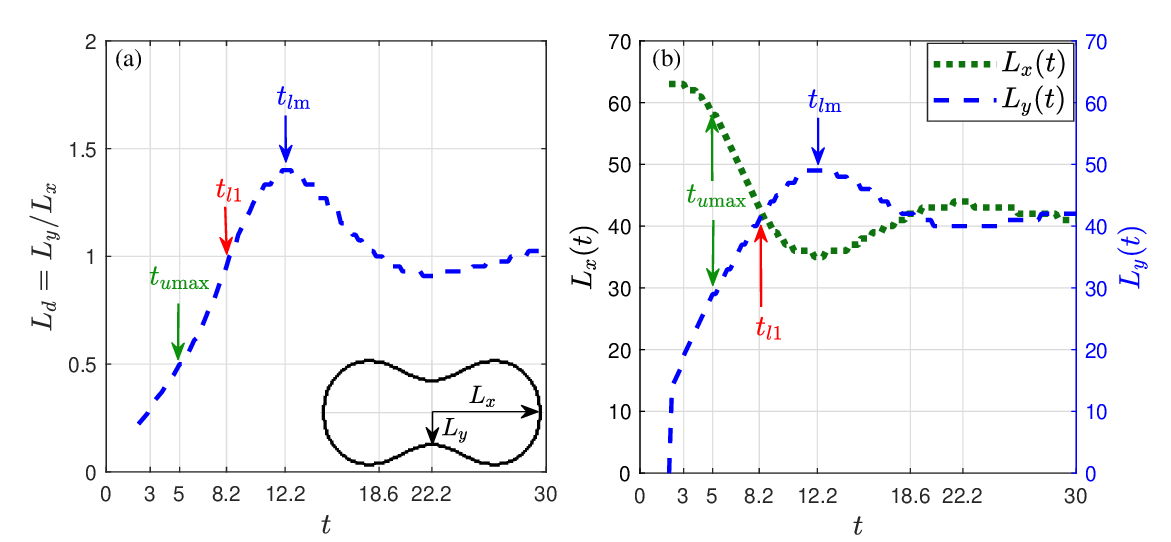}}
 \caption{\centering{Temporal evolutions of (a) the ratio of minor and major axes $L_d = L_y/L_x$ and (b) the major axis $L_x$ and the minor axis $L_y$.}}
 \label{F2}
\end{figure*}
Figure \ref{F1} depicts the vector diagram of the flow field at eight characteristic moments during the coalescence of two stationary droplets. The solid blue line within the figure delineates the boundary contour between the vapor and liquid phases. Similar to the merging process of two initial static bubbles, the coalescence of two equal-sized static droplets also undergoes the rapid coalescence stage characterized by morphological changes such as dumbbell-shaped and spindle-shaped configurations, followed by the initial unsteady circular form, succeeded by a damping oscillation phase. Furthermore, four small vortices initially form between the two droplets during the coalescing process. The size, area and direction of the vortex velocity field vary as the coalescing process. As depicted in Fig. \ref{F2}, three distinct moments exist during the merging process. The first one is the moment of the ratio of minor and major axes $L_d = 0.5$, where $L_d = L_y/L_x$ with $L_x$ and $L_y$ being the major axis and minor axis of the new droplet, respectively. At this instant, $L_y\approx r_{0\rm l}= r_{0\rm r}$ and the merging speed reaches its maximum because the molecules inside the droplet are fully driven. Therefore, this moment is marked as $t_{u\rm max}$. The second moment occurs at $L_d = 1.0$ when the new droplet first forms an unstable circle, and it is labeled as $t_{l1}$. The third moment, denoted as $t_{l\rm m}$, occurs  when $L_d$ reaches its maximum.
\begin{figure*}[htbp]
{\centering
\includegraphics[width=0.8\textwidth]{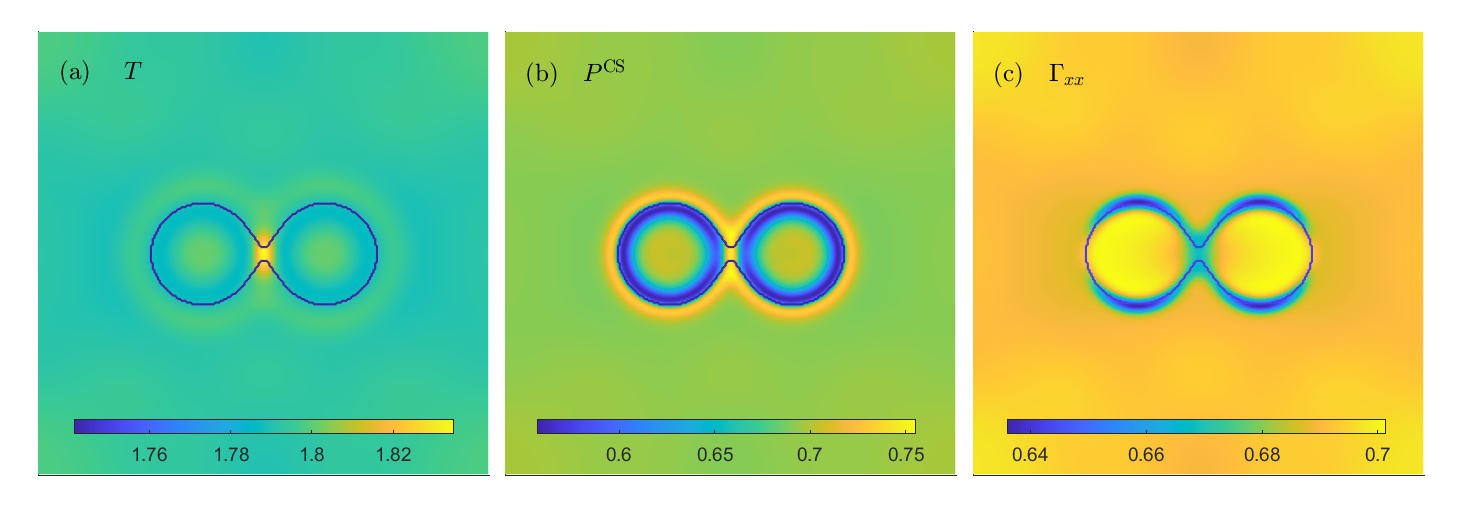}}
 \caption{\centering{Spatial distribution maps of temperature $T$, hot pressure ${P^{{ \rm {CS}}}} $, and the $xx$ component of the total pressure tensor ${\Gamma_{xx}} $ at $t= 1.0$. Here, $K = 0.0002$, $\tau  = 0.0025$,  $\Pr  = 0.5$, ${r_{0{\rm{l}}}} = {r_{0{\rm{r}}}} = 30$, and each submap covers the entire computational domain  ${N_x} \times {N_y} = 256 \times 256$.}}
 \label{F3}
\end{figure*}

When the distance between two stationary droplets is of the same order as the width of the liquid-vapor boundary layer, they tend to approach each other driven by the force of the total pressure gradient. As they get closer, a liquid bridge begins to form at their initial point of contact. This region experiences a substantial decrease in the total pressure tensor ${\bm{\Gamma }} = {P^{{\rm{CS}}}}{\bf{I}} + {\bm{\Lambda }}$ due to the pronounced negative curvature near the contact point, as illustrated in Fig. \ref{F3}(c). In the early stage of droplet coalescence, the temperature in the center area of the liquid bridge is higher [see Fig. \ref{F3}(a)], resulting in a higher thermal pressure according to the EOS, as shown in Fig. \ref{F3}(b). Therefore, the pressure gradient contributed by $\bm{\Lambda }$ becomes the primary driving force that promotes droplet coalescence.
 \begin{figure*}[htbp]
{\centering
\includegraphics[width=0.8\textwidth]{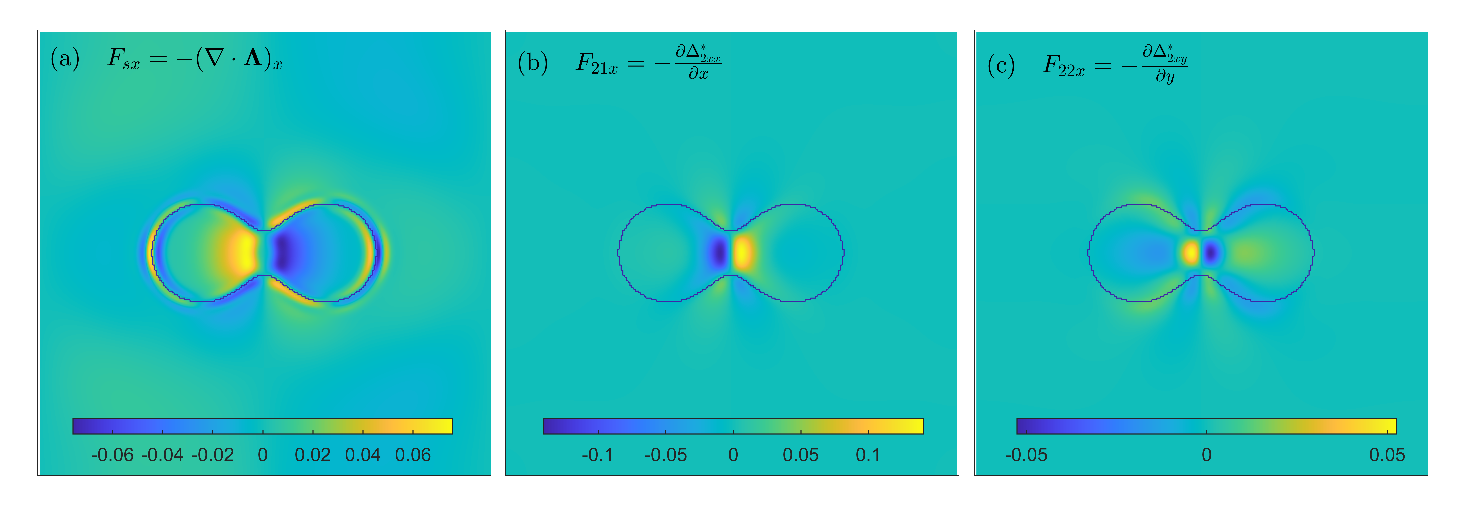}}
 \caption{\centering{Spatial distributions of ${F_{sx}} = {( - \bm{\nabla}  \cdot {\bm{\Lambda }})_x}$, ${F_{2x1}} = {{ - \partial_x} \Delta _{2xx}^*}$ and ${F_{2x2}} =  - {{\partial_y} \Delta _{2xy}^*}$ at $t=2.0$. In each submap, the grid covers the entire computational domain ${N_x} \times {N_y} = 256 \times 256$, and the color bar at the bottom of each panel indicates the magnitude.}}
 \label{F4}
\end{figure*}
\begin{figure*}[htbp]
{\centering
\includegraphics[width=0.8\textwidth]{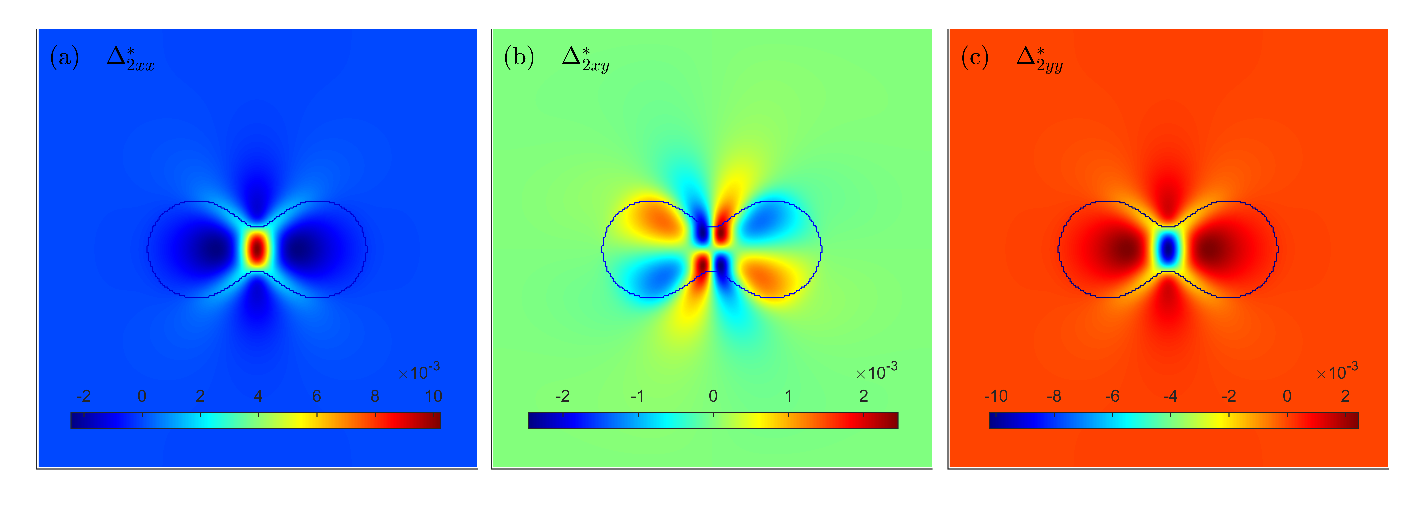}}
 \caption{\centering{Spatial distributions of $\Delta _{2xx}^*$ (a), $%
\Delta _{2xy}^*$ (b), and $\Delta _{2yy}^*$ (c), at $t=2.0$. In each submap, the grid covers the entire computational domain computational domain ${N_x} \times {N_y} = 256 \times 256$, and the color bar at the bottom of each panel indicates the magnitude of $\Delta_{2\alpha \beta}^*$.}}
\label{F5}
\end{figure*}

In equation (5), the force ${{\bf{F}}_s} =  - \bm{\nabla}  \cdot {\bm{\Lambda }}$ represents the contribution of the surface tension, while the force ${{\bf{F}}_2} =  - \bm{\nabla}  \cdot {\bm{\Delta }}_2^*$ corresponds to the contribution of the NOMFs. Their components in the merging direction ($x$-direction) are of particular interest. Specifically, their components in the $x$-direction are denoted as ${F_{sx}} = {( - \bm{\nabla}  \cdot {\bm{\Lambda }})_x}$ and ${F_{2x}} = {( - \bm{\nabla}  \cdot {\bm{\Delta }}_2^*)_x} =  - {{\partial_x} \Delta _{2xx}^*} - {{\partial_y} \Delta _{2xy}^*}$, displayed in Fig. \ref{F4} for moment at $t=2.0$. As shown in Fig. \ref{F4}(a), ${F_{sx}} = {( - \bm{\nabla}  \cdot {\bm{\Lambda }})_x}$ exhibits positive near the left side of the liquid bridge and negative near the right side of the liquid bridge. This indicates that the gradient force on the left side points to the right, whereas on the right side, it points to the left. As previously discussed, surface tension serves as the primary driving force for droplet coalescence.

A comparison between Fig. \ref{F4}(b) and Fig. \ref{F4}(c) reveals two distinct forces attributed to ${\bm{\Delta }}_2^*$. Specifically, the force ${F_{2x1}} = {{ - \partial_x} \Delta _{2xx}^*}$, originating from $\Delta _{2xx}^*$, is identified as a driving force, while the force ${F_{2x2}} =  - {{\partial_y} \Delta _{2xy}^*}$ provided by $\Delta _{2xy}^*$ acts as a resisting force. It can be inferred from  $\bm{\Delta }_{2}^{\ast (1)}=-\rho T\tau [{\bm {\nabla} {\bf u}+{{(\bm {\nabla} {\bf u})}^{Tr}}-\mathbf{I}\bm {\nabla} \cdot \mathbf{u}}]$ that the spatial distribution of ${\bm{\Delta }}_2^*$ is determined by the distribution of velocity (see Fig. \ref{F1}). Further analysis reveals that $\Delta _{2xx}^*$ is dominated by ${\partial_x} {u_x}$ and $\Delta _{2xy}^*$ is dominated by ${\partial_x} {u_y}$. Consequently, $\Delta_{2xx}^*$ displays positive values in the central region and negative values on both sides, while $\Delta _{2xy}^*$ manifests an anti-symmetric double quadrupole structure at $t=2.0$, as depicted in Fig \ref{F5}. More importantly, during the early stage of droplet coalescence, the surface tension and capillary effects are the main driving factors for increasing the radius of the liquid bridge and velocity perpendicular to the merging direction $u_y$\cite{RN201,RN202}, while $u_x$ increases passively with the increase of the radius of the liquid bridge. Therefore, ${F_{2x1}} = {{ - \partial_x} \Delta _{2xx}^*}$ inhibits the droplet coalescence process, whereas ${F_{2x2}} =  - {{\partial_y} \Delta _{2xy}^*}$ promotes it. On the whole,  ${F_{2x}} = {( - \bm{\nabla}  \cdot {\bm{\Delta }}_2^*)_x}$ suppresses droplet coalescence, because the maximum value of ${F_{2x1}}$ is approximately 2.6 times that of ${F_{2x2}}$, i.e. $({{ - \partial_x} \Delta _{2xx}^*})_{\max } \approx 2.6{( - {{\partial_y} \Delta _{2xy}^*})}_{\max }$ in the considered case.
\begin{figure}[htbp]
{\centering
\includegraphics[width=0.43\textwidth]{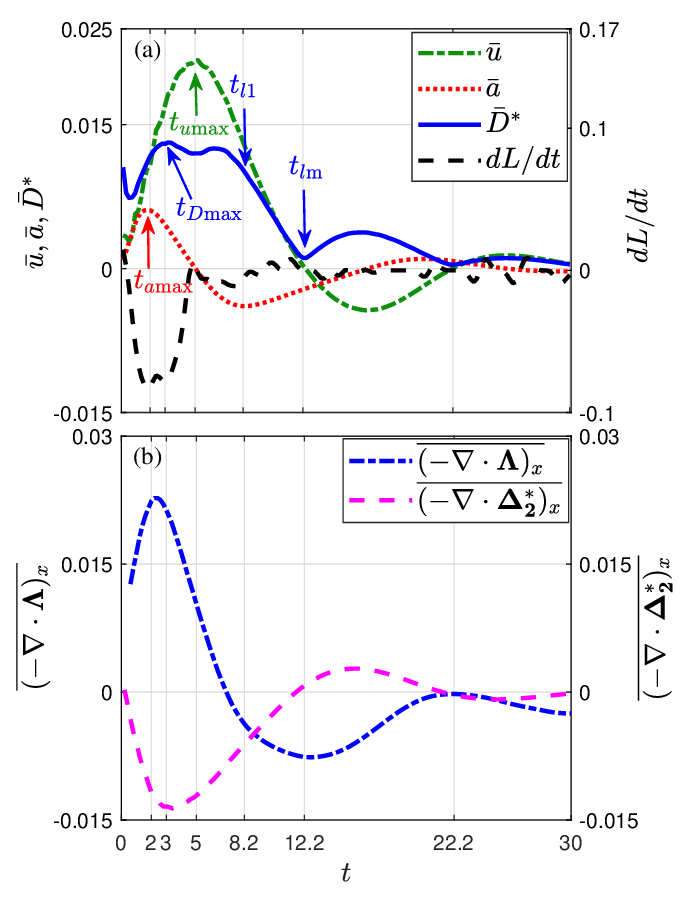}}
 \caption{\centering{(a) Temporal profiles of the speed of droplet coalescence $\bar u(t)$, the acceleration of droplet coalescence $\bar a(t) = d{\bar u} / {dt}$, the strength of total non-equilibrium ${\bar D^*}$,and the decreasing rate of boundary length $d{L}/{dt}$; (b) Time evolutions of ${\bar F_{sx}} = \overline {{{( -\bm{\nabla}  \cdot {\bm{\Lambda }})}_x}} (t)$ and  ${\bar F_{2x}} = \overline {{{( - \bm{\nabla}  \cdot {\bm{\Delta }}_2^*)}_x}} (t)$.}}
 \label{F6}
\end{figure}
\begin{figure*}[htbp]
{\centering
\includegraphics[width=0.8\textwidth]{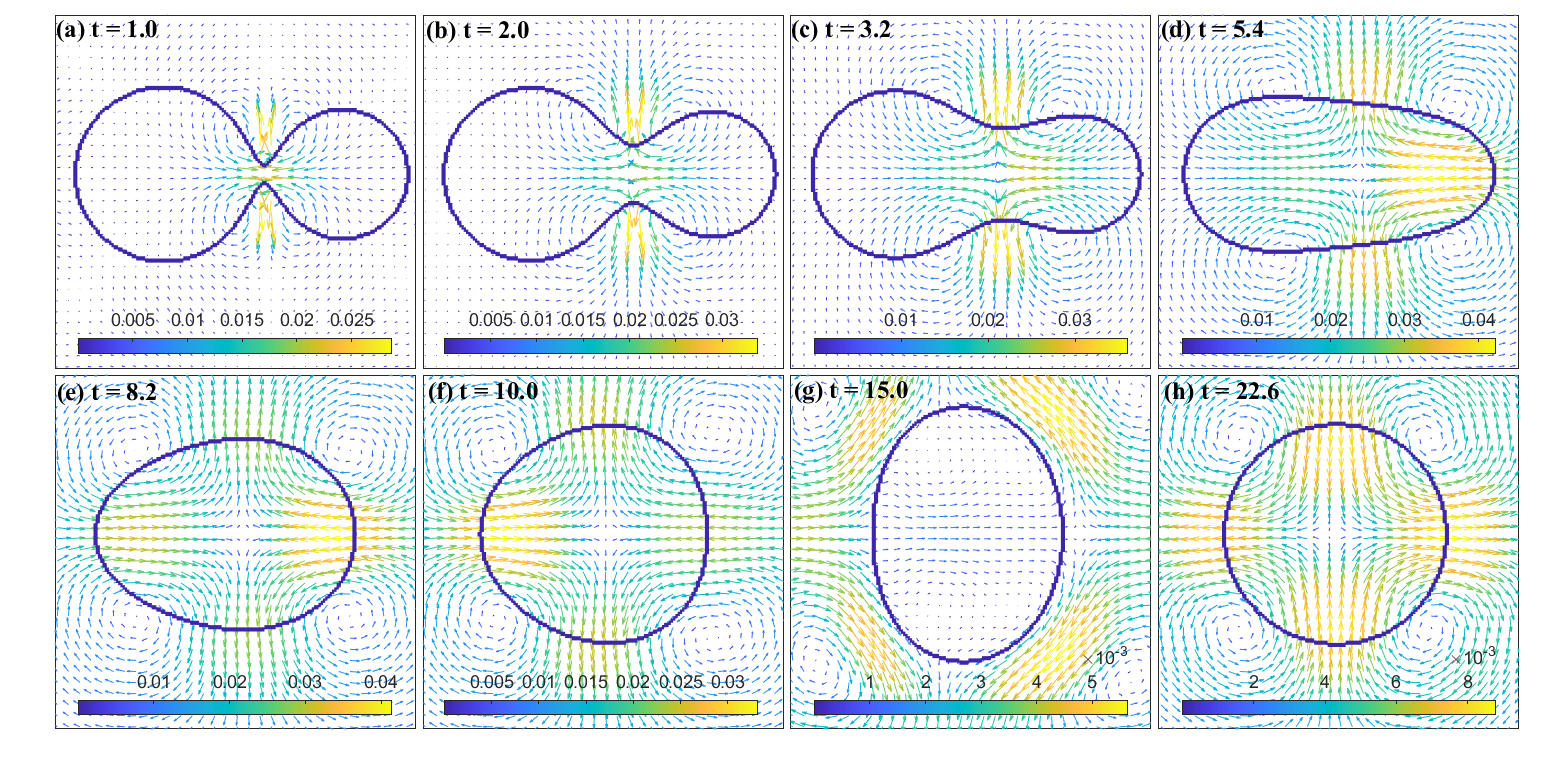}}
 \caption{\centering{Spatial distributions of velocity vector at $t=1.0$, $2.0$, $3.2$, $5.4$, $8.2$, $10.0$, $15.0$ and $22.6$ during the merging process of the two unequal droplets. Here, ${r_{0{\rm{l}}}} = 40,  {r_{0{\rm{r}}}} = 30$, and the grid range of each subpicture is the partial simulated area ${N'_x} \times {N'_y} = 150 \times 150$.}}
 \label{F7}
\end{figure*}
\begin{figure*}[htbp]
{\centering
\includegraphics[width=0.8\textwidth]{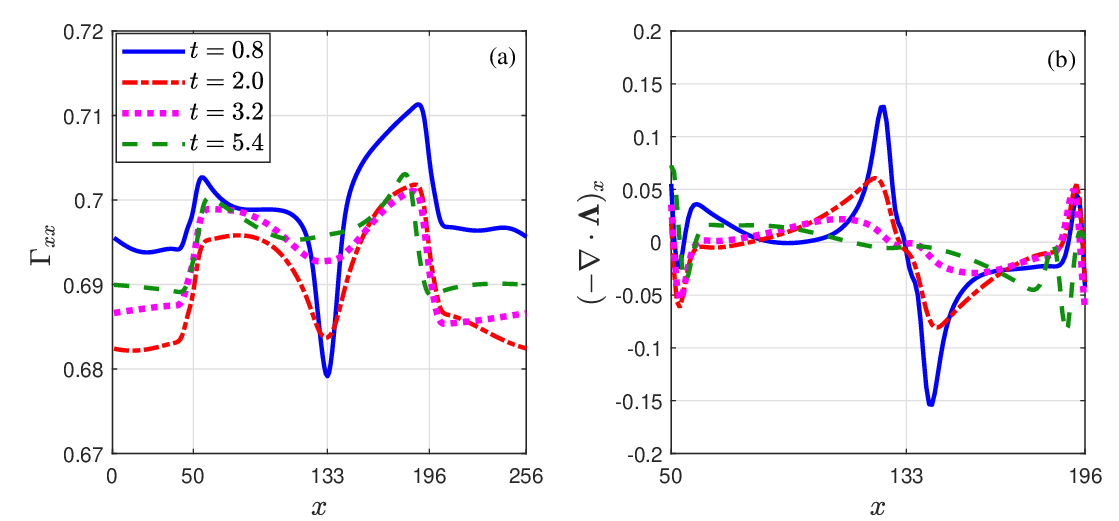}}
 \caption{\centering{Spatial distributions of the $xx$ component of the total pressure tensor ${\Gamma_{xx}} $ and ${F_{sx}} = {( - \bm{\nabla}  \cdot {\bm{\Lambda }})_x}$ along the horizontal central axis of two droplets at $t=0.8$, $2.0$, $3.2$, and $5.4$. Here, $x=50$ and $x=196$ correspond to the left and right coordinates of the liquid-vapor boundary at $t=0.8$.}}
 \label{F8}
\end{figure*}

To further study the relationship between the strength of total non-equilibrium and the morphological and dynamic characteristics of droplet coalescence, we analyze the time evolutions of the speed of droplet coalescence $\bar u(t)$, acceleration $\bar a(t) = d{{\bar u}/{dt}}$, boundary length reduction rate ${dL} /{dt}$,  the total TNE strength ${\bar D^*}(t)$, spatial statistics averages of ${ F_{sx}}$,  ${\bar F_{sx}} = \overline {{{( -\bm{\nabla}  \cdot {\bm{\Lambda }})}_x}}(t)$, and the mean of spatial statistics of ${F_{2x}}$, ${\bar F_{2x}} = \overline {{{( - \bm{\nabla}  \cdot {\bm{\Delta }}_2^*)}_x}}(t)$. Their definitions are as follows:
 \begin{equation}
\bar{u}(t)={\frac{{\sum {\rho (x,y,t){u_{x}}(x,y,t)}}}{{\sum {\rho
(x,y,t)}}}},
\end{equation}
\begin{equation}
\bar D^*(t)={\frac{{\sum {\rho (x,y,t)\sqrt{\bm{\Delta }%
_{2}^{\ast 2}+\bm{\Delta }_{3}^{\ast 2}+\bm{\Delta }_{3,1}^{\ast 2}+ \bm{\Delta }_{4,2}^{\ast 2}}}}}{{\sum {\rho (x,y,t)}}}},
\end{equation}
\begin{equation}
  {\bar F_{sx}} = \overline {{{( - \bm{\nabla}  \cdot {\bm{\Lambda }})}_x}} (t) = {{\sum {\rho (x,y,t){{( - \bm{\nabla}  \cdot {\bm{\Lambda }})}_x}} } \over {\sum {\rho (x,y,t)} }},
\end{equation}
\begin{equation}
   {\bar F_{2x}} = \overline {{{( - \bm{\nabla}  \cdot {\bm{\Delta }}_2^*)}_x}} (t) =  - {{\sum {\rho (x,y,t){{( - \bm{\nabla}  \cdot {\bm{\Delta }}_2^*)}_x}} } \over {\sum {\rho (x,y,t)} }}.
\end{equation}
As demonstrated by Fig. \ref{F6}(a), before the merging speed reaches its maximum ($t \le {t_{u {\max}}} = 5.0$), the length of the liquid-vapor boundary experiences a rapid reduction, with its rate of decrease peaking at $2.0 \leq t \leq 3.0$. Simultaneously, the acceleration of droplet coalescence $\bar a(t) = d{\bar u}/dt$
and the strength of total TNE ${\bar D^*}$ both reach their peaks successively (${t_{a{\max}}} = 2.0$ and ${t_{D{\max} }} = 3.0$), owing to the maximum releasing rate of  surface energy. In Fig. \ref{F6}(b), the driving force, ${\bar{F}_{sx}} = \overline{(-\bm{\nabla} \cdot \bm{\Lambda})}_x(t)$, attains its maximum at the same time as $\bar{a}(t)$ (at $t = {t_{{a{\max}}}} = 2.0$), while the resistive force, ${\bar F_{2x}} = \overline{(-\bm{\nabla} \cdot \bm{\Delta}_2^*)}_x(t)$, reaches its peak at $t = {t_{D{{\max}}}} = 3.0$.
 Furthermore, it's important to note that $\left| {{{\bar F}_{sx}} = \overline {{{( - \nabla  \cdot {\bm{\Lambda }})}_x}} } \right| > \left| {{{\bar F}_{2x}} = \overline {{{( - \nabla  \cdot {\bm{\Delta }}_2^*)}_x}} } \right|$, confirming the occurrence of droplet coalescence.

By comprehensively analyzing the time evolution curves of $L_d$, $\bar u$ and $\bar D^*$, the time interval of $t \le t_{u \rm max}$ can be defined as the rapid coalescing stage of two equal-size droplets, followed by a damped oscillation stage. Importantly, except for the first valley value of $\bar D^*$ at $t = {t_{u {\max}}}$, each instant that $\bar D^*$ reaches a valley corresponds to an extremum in $L_d$. This phenomenon can be attributed to three factors: (i) the total TNE strength of this system is mainly determined by the strength of NOMFs (${\bm{\Delta }}^*_2$) and fluxes of heat flux (${\bm{\Delta }}^*_{4,2}$); (ii) under unchanged parameters, the time evolutions of ${\bm{\Delta }}^*_2$ and ${\bm{\Delta }}^*_{4,2}$ are mainly determined by the spatiotemporal evolution of velocity gradient \cite { RN30, RN107}; (iii) when $L_d$ reaches an extreme value, the droplet possesses either its widest shape in the $x$-direction or its tallest shape in the $y$-direction [see Fig. \ref{F1}(e) and  Fig. \ref{F1}(g)], and at this time, the velocity gradient within the droplet is minimized in the current oscillation period. Additionally, for the time interval $t_{u {\max}}< t < t_{l1}$, there exists a peak on the $\bar D^*$ evolution profile due to an increase in velocity gradient within the droplet. Essentially, this occurs because particles within the droplet are further propelled forward as $L_x$ rapidly decreases, and the decay rate of $L_x$ surpasses the growth rate of $L_y$ during $t_{u {\max}}< t < t_{l1}$ [see the green dots line in Fig. \ref{F2}(b)].

In summary, the temporal evolution of the total TNE intensity provides a significative, convenient and efficient physical criterion for distinguishing the stages of droplet coalescence, which deepens our understanding of droplet merging mechanism from the perspective of kinetics.
\subsection{Non-equilibrium characteristics during the coalescence of two unequal-size droplets}
The coalescence of droplet of unequal size is more common in natural occurrences and engineering
scenarios than the coalescence of droplets with equal size.
 A noteworthy aspect of the coalescence process involving two initially static droplets with differing sizes is the absorption of the smaller droplet by the larger one. This results in the amalgamation of the two droplets, ultimately forming a larger one. Subsequently, this merged droplet experiences a sequence of damping and oscillation events before eventually reaching a stable equilibrium state. In this section, we embark on a comprehensive analysis of the intricate physical mechanisms and the interplay of non-equilibrium effects that govern the coalescence process of two unequal droplets. The parameters used in the numerical calculations are $K = 0.0002$, $\tau  = 0.0025$, $\Pr  = 0.5$.

As displayed in Figs. \ref{F7}(a)-(d), the two droplets are in close proximity to each other due to the influence of surface tension. Notably, the smaller droplet on the right-hand side experiences a pronounced initial increase in velocity immediately after the two droplets come into contact. The reason is that in the early stage of coalescence, the total pressure inside the smaller droplet is stronger due to its larger curvature, as demonstrated in Fig. \ref{F8}. Consequently, the total pressure, pressure gradient and the absolute value of ${F_{sx}} = {( - \bm{\nabla}  \cdot {\bm{\Lambda }})_x}$ within the smaller droplet on the right are all greater than those in the larger droplet on the left. After entering the damping oscillation stage, there is also a period of asymmetric oscillation [see Figs. \ref{F7}(f)-(h)].
\begin{figure}[htbp]
{\centering
\includegraphics[width=0.46\textwidth]{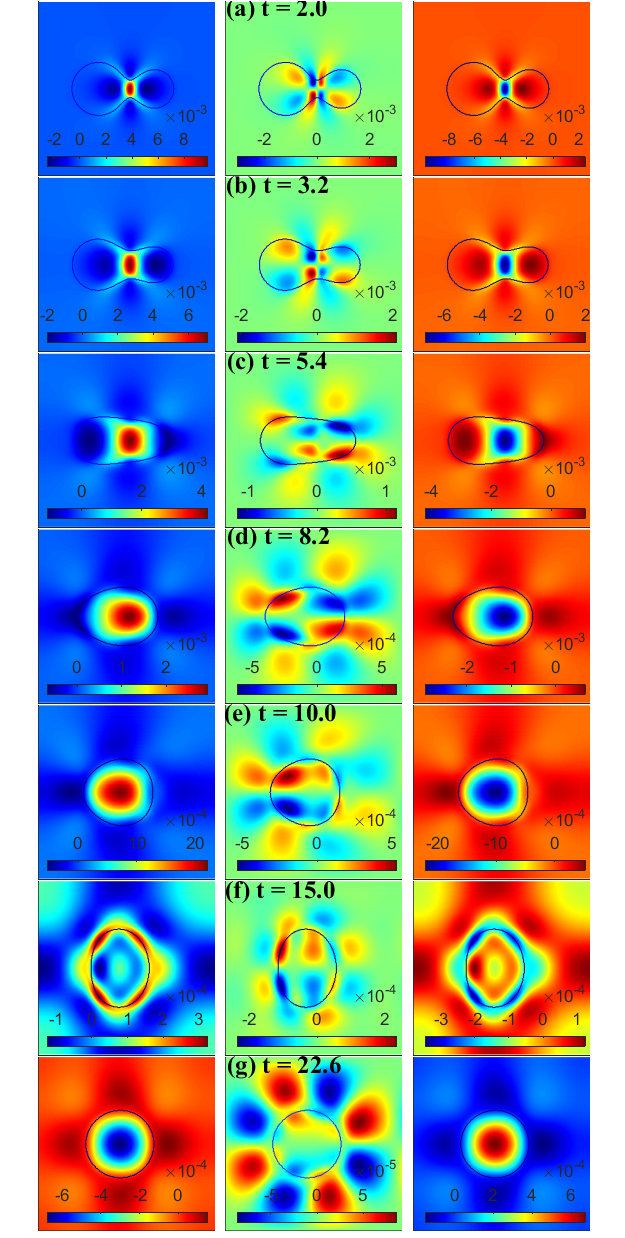}}
 \caption{\centering{Spatial distributions of $\Delta _{2xx}^*$ (first column), $\Delta _{2xy}^*$ (second column), and $\Delta _{2yy}^*$ (third column). In each submap, the grid covers the entire computational domain ${N_x} \times {N_y} = 256 \times 256$, and the color bar at the bottom of each panel indicates the magnitude of $\Delta_{2\alpha \beta}^*$.}}
 \label{F9}
\end{figure}
\begin{figure}[htbp]
{\centering
\includegraphics[width=0.4\textwidth]{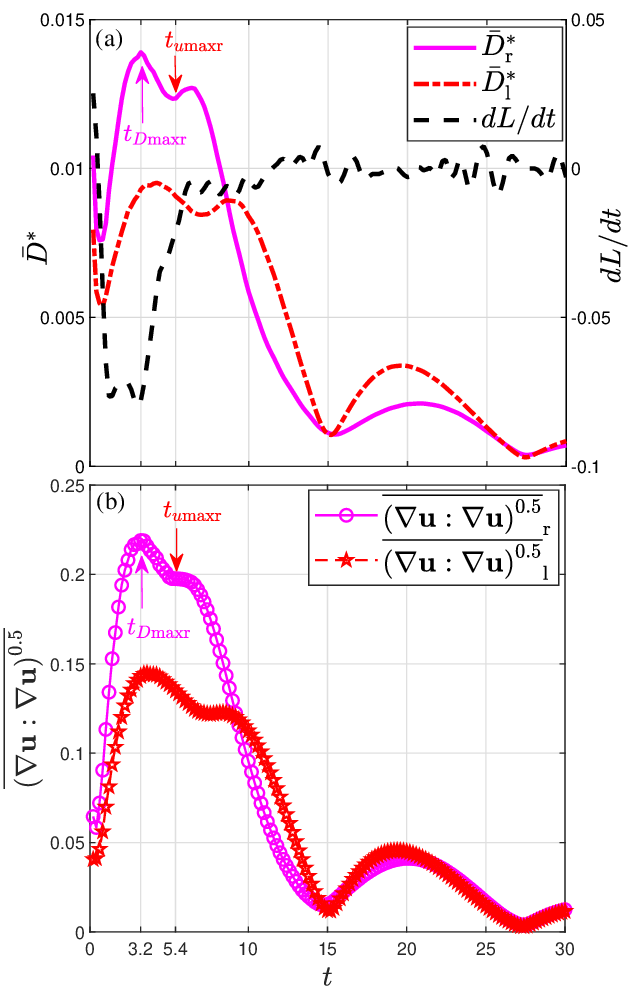}}
 \caption{\centering{(a) The temporal profiles of $\bar D^*_{\rm r}$(the strength of total TNE in the right droplet), $\bar D^*_{\rm l}$(the strength of total TNE in the left droplet), and the slope of boundary length $dL/dt$; (b) the spatial average of the velocity gradient $\overline {{{(\bm{\nabla} {\bf{u}}:\bm{\nabla} {\bf{u}})}^{0.5}}}$ in the right and left deoplets.}}
 \label{F10}
\end{figure}
\begin{figure}[htbp]
{\centering
\includegraphics[width=0.4\textwidth]{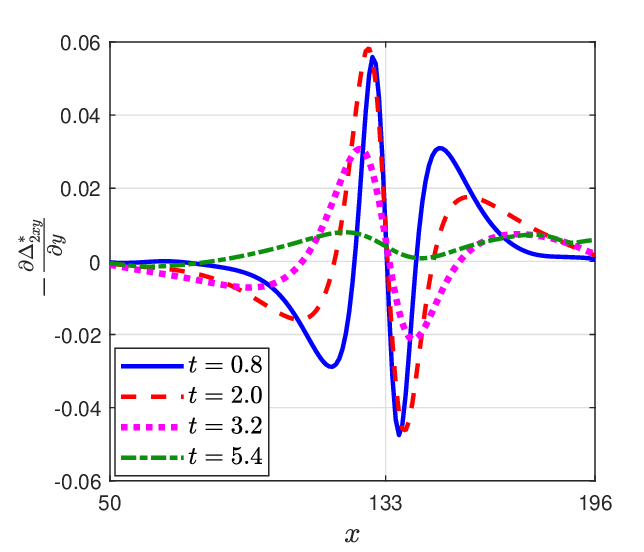}}
 \caption{\centering{Distributions of ${F_{2x2}} =  - \partial \Delta _{2xy}^* / {\partial y}$ along the horizontal central axis  of two droplets at $t=0.8$, $2.0$, $3.2$, and $5.4$. Here, $x=50$ and $x=196$ represent the left and right coordinates of the liquid-vapor boundary at $t=0.8$.}}
 \label{F11}
\end{figure}
\begin{figure*}[htbp]
{\centering
\includegraphics[width=1.\textwidth]{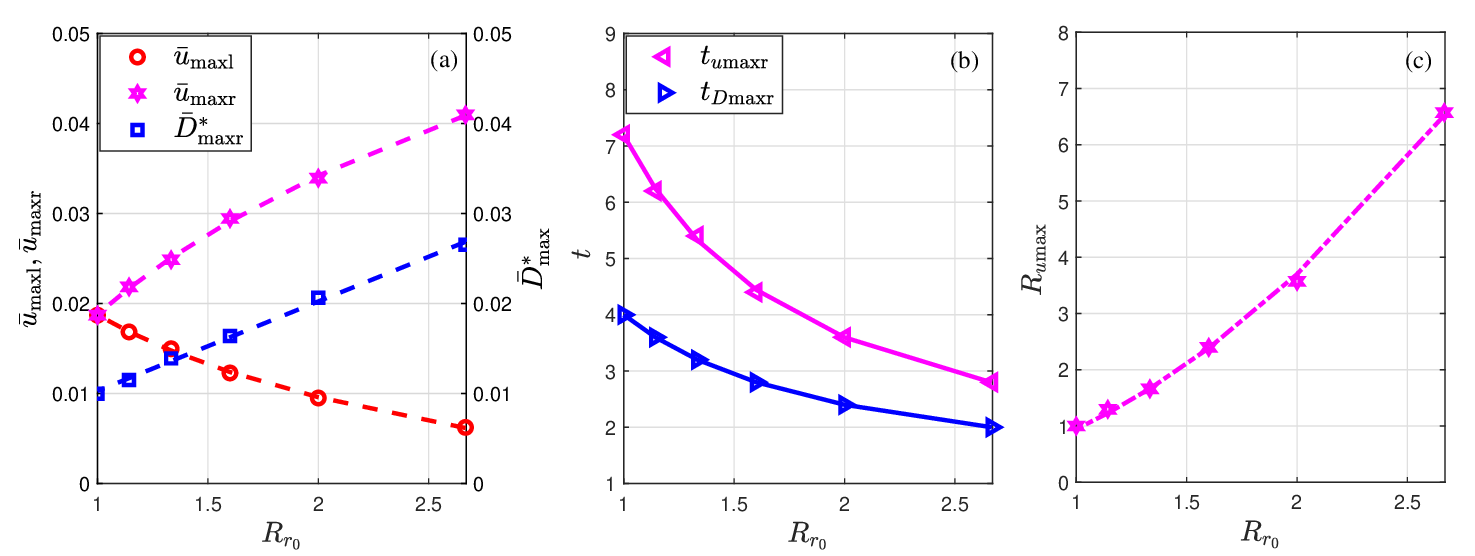}}
 \caption{\centering{(a) Influences of the initial radius ratio ${R_{{r_0}}} = {r_{0 \rm l}} / {r_{0\rm r}}$ on the the maximum of mean coalescence velocity in left droplet ${\bar u_{{\rm{maxl}}}}$, the the maximum of mean coalescence velocity in right droplet ${\bar u_{{\rm{maxr}}}}$, and the maximum of total TNE strength in the right droplet $\bar D^*_{{\mathop{\rm maxr}\nolimits} }$; (b) the effects of  ${R_{{r_0}}}$ on the instant when the mean coalescence velocity in the right droplet reaches the maximum ${t_{{u}{\rm{maxr}}}}$, and the instant when $\bar D^*$
in the right droplet reaches the maximum ${t_{{D}{\rm{maxr}}}}$; (c) Relation between the
maximum of the mean coalescence velocity in the small and large droplets ${R_{u\rm max}} = {\bar u}_{\rm{maxr}}/{{\bar u}_{\rm{maxl}}}$ and ${R_{{r_0}}}$.}}
\label{F12}
\end{figure*}
 \begin{figure*}[htbp]
{\centering
\includegraphics[width=0.7\textwidth]{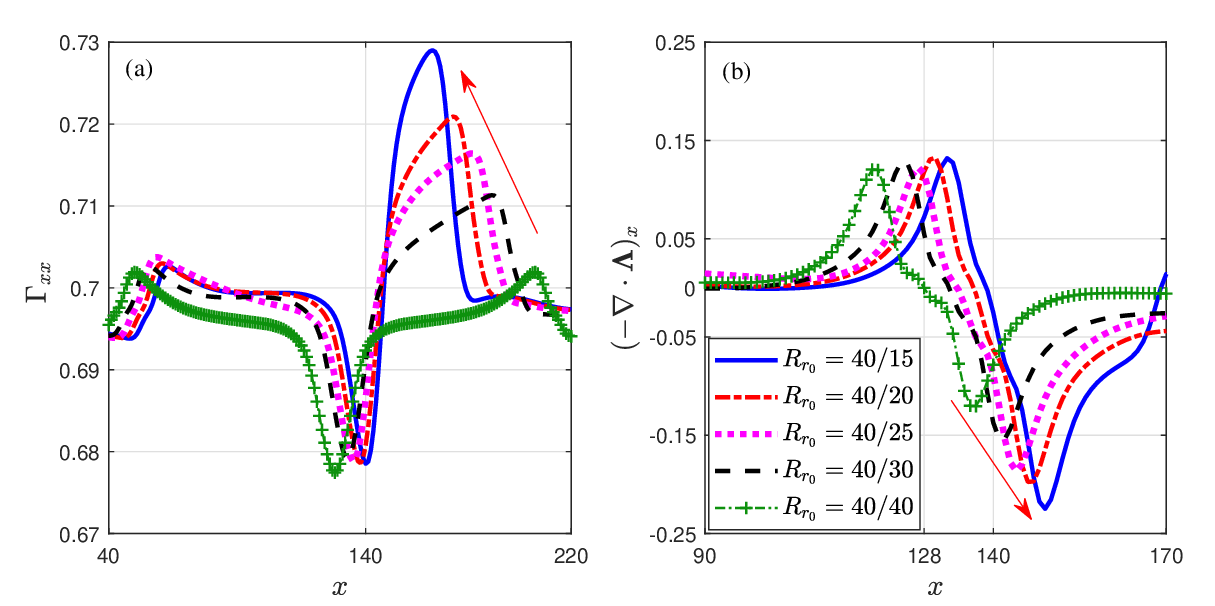}}
 \caption{\centering{Spatial distributions of the $xx$ component of the total pressure tensor ${\Gamma_{xx}} $ and ${F_{sx}} = {( - \bm{\nabla}  \cdot {\bm{\Lambda }})_x}$ along the horizontal central axis of two droplets for various ${R_{{r_0}}} = {r_{0 \rm l}} / {r_{0\rm r}}$ at $t=0.8$ when the absolute values of ${\Gamma_{xx}} $ and ${F_{sx}}$ are both at their maximum.}}
 \label{F13}
\end{figure*}
\begin{figure*}[htbp]
{\centering
\includegraphics[width=0.9\textwidth]{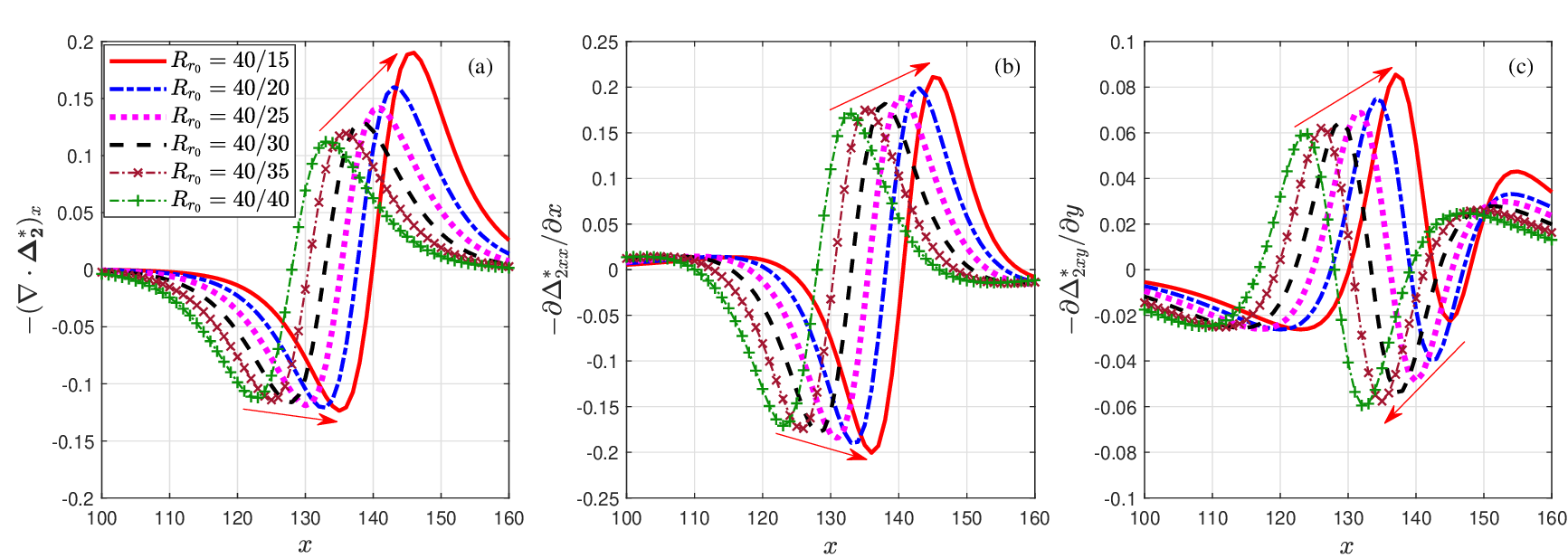}}
 \caption{\centering{Curves of  ${F_{2x}} = {( - \bm{\nabla}  \cdot {\bm{\Delta }}_2^*)_x} =  - {{\partial \Delta _{2xx}^*}/ {\partial x}} - {{\partial \Delta _{2xy}^*} / {\partial y}}$, ${F_{2x1}} = { - \partial \Delta _{2xx}^*} / {\partial x}$, and ${F_{2x2}} =  -{ \partial \Delta _{2xy}^*}/{\partial y}$ on the horizontal central axis  of two droplets for various initial radius ratios at $t=1.2$  when the absolute values of ${F_{2x1}} $ and ${F_{2x2}}$ are both at their maximum.}}
 \label{F14}
\end{figure*}

The asymmetry of velocity distribution and velocity gradient within the droplets inevitably results in an asymmetrical spatial distribution of the TNE effects. When $t<5.4$, the amplitudes of $\Delta_{2xx }^*$ and $\Delta_{2yy }^*$ inside the smaller droplet are greater than those inside the larger droplet, as depicted in the first and third columns of Fig. \ref{F9}(a) and Fig. \ref{F9}(b).  The red dot-dash line and the purple solid line in Fig. \ref{F10}(a) show the time evolution curves of the total TNE strength $\bar D^*$ in the left larger droplet and the right smaller droplet, respectively.  Obviously, the total TNE strength in the smaller droplet is stronger than that in the larger droplet, and they both reach their peak values when the liquid-vapor boundary length experiences the most rapid decrease. This is because the average velocity spatial gradient within the smaller  $\overline {{{(\bm{\nabla} {\bf{u}}:\bm{\nabla} {\bf{u}})}^{0.5}}}$ [$={\sum {\rho (x,y,t){{(\nabla {\bf{u}}:\nabla {\bf{u}})}^{0.5}}} }/{\sum {\rho (x,y,t)} }$] is higher than that in the larger droplet and $\overline {{{(\bm{\nabla} {\bf{u}}:\bm{\nabla} {\bf{u}})}^{0.5}}}$ also reaches its peak when $L$ decreases most rapidly [see Fig. \ref{F10}(b)]. In particular, although the total TNE strength in the smaller droplet is stronger in the early stage of the coalescence, the non-equilibrium effect related to the shear velocity, such as $\Delta_{2xy}^*$, is stronger in the larger droplet [see the middle column in Fig. \ref{F9}(a) and Fig. \ref{F9}(b)] primarily due to the stronger shear velocity within the larger droplet. During the early phase of droplet coalescence, the absolute value of ${F_{2x2}} =  - {{\partial_y} \Delta _{2xy}^*}$ in the left adjacent region of the liquid bridge is also greater than that in the right adjacent region of the liquid bridge, as exhibited in Fig. \ref{F11}.
 \begin{figure}[htbp]
{\centering
\includegraphics[width=0.5\textwidth]{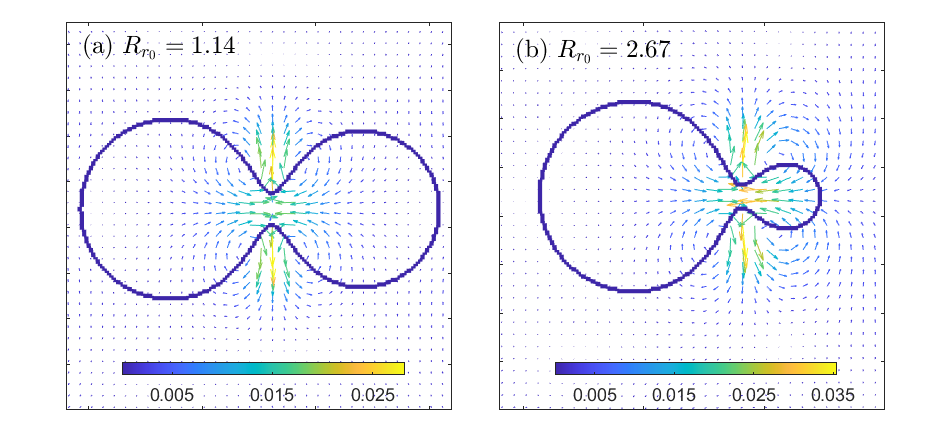}}
 \caption{\centering{Spatial distributions of velocity vectors for ${R_{{r_0}}} = 40/35$ and ${R_{{r_0}}} = 40/15$ at $t=1.2$.}}
 \label{F15}
\end{figure}

Figure \ref{F12} presents the effects of the initial radius ratio of the left and right droplets, denoted as ${R_{{r_0}}} = {{{r_{0{\rm{l}}}}}/ {{r_{0{\rm{r}}}}}}$, on various parameters including ${\bar u_{{\rm{maxl}}}}$ (the maximum of mean coalescence velocity in left larger droplet),  ${\bar u_{{\rm{maxr}}}}$ (the maximum of mean coalescence velocity in the right smaller droplet), $\bar D_{{\mathop{\rm maxr}\nolimits} }^*$ (the maximum of the total TNE strength $\bar D^*$ in the right smaller droplet), ${t_{{u}{\rm{maxr}}}}$ (the instant when the mean coalescence velocity in the right smaller droplet reaches the maximum), ${t_{{D}{\rm{maxr}}}}$ (the instant when the total TNE strength  $\bar D^*$ in the right smaller droplet reaches the maximum), and ${R_{{u}{\max}}} =  {{\bar u}_{\rm{maxr}}} / {{\bar u}_{\rm{maxl}}}$ (the ratio of the maximum of the mean coalescence velocity in the left larger and right smaller droplets). Figure \ref{F12}(a) displays the functional relationships between ${\bar u_{{\rm{maxl}}}}$, ${\bar u_{{\rm{maxr}}}}$, $\bar D_{{\mathop{\rm maxr}\nolimits} }^*$ and ${R_{{r_0}}}$ , which can be described as follows:  ${\bar u_{{\rm{maxl}}}} = 0.068R_{{r_0}}^{ - 1/5} - 0.049$, ${\bar u_{{\rm{maxr}}}} = 0.23R_{{r_0}}^{0.093} - 0.21$, and $\bar D_{{\mathop{\rm maxr}\nolimits} }^* = 0.0099{R_{{r_0}}} + 0.00034$. Figure \ref{F12}(b) presents the functional relationships between ${t_{{u}{\rm{maxr}}}}$, ${t_{{D}{\rm{maxr}}}}$ and ${R_{{r_0}}}$, which are represented as follows ${t_{{u}{\rm{maxr}}}} = 6.3R_{{r_0}}^{ - 6/5} + 0.88$ and ${t_{{D}{\rm{maxr}}}} = 3.2R_{{r_0}}^{ - 1} + 0.80$. With the increase of ${R_{{r_0}}}$, the absolute values of ${\Gamma _{xx}}$  and ${( - \bm{\nabla} \cdot {\bm{\Lambda }})_x}$  increase gradually in the smaller droplet, while they change very weakly in the large droplet (see Fig. \ref{F13}). Thus, both ${\bar u_{{\rm{maxr}}}}$ and $\bar D_{{\mathop{\rm maxr}\nolimits} }^*$  increase as ${R_{{r_0}}}$ increases, whereas both ${t_{{u}{\rm{maxr}}}}$ and ${t_{{D}{\rm{maxr}}}}$ decrease with the increase of  ${R_{{r_0}}}$. A reduction in the radius of the right smaller droplet results in a decrease in the liquid bridge radius.
Consequently, the reduced radius of the small droplet on the right side leads to a decrease in the number of molecules entering the large droplet on the left side. Simultaneously, the impact resistance of the right molecule to the left molecule gradually increases with the rising mean coalescence velocity in the small droplet.
Therefore, the maximum value of the mean coalescence velocity ${\bar u_{{\rm{maxl}}}}$ in the left big droplet gradually decreases as ${R_{{r_0}}}$ increases. In addition, the relationship between the ratio of the maximum coalescence velocity of the small and large droplets ${R_{{u}{\max} }} = {{\bar u}_{\rm{maxr}}} / {{\bar u}_{\rm{maxl}}}$ follows a quadratic function ${R_{{u}\rm max }} \sim R_{{r_0}}^2$ (specifically, ${R_{{u}\rm max }} = 0.94R_{{r_0}}^{1.97}$).

During the coalescence of two unequal droplets, the spatial distribution of  ${F_{2x}} = {( - \bm{\nabla}  \cdot {\bm{\Delta }}_2^*)_x} =  - {{\partial_x} \Delta _{2xx}^*} - {{\partial_y} \Delta _{2xy}^*}$ also displays an asymmetry. As illustrated by Fig. 14(a), with the increase of ${R_{{r_0}}}$, the absolute values of ${F_{2x}} = {( - \bm{\nabla}  \cdot {\bm{\Delta }}_2^*)_x}$ in the left and right droplets increase. This increase signifies an increase in the resistant force contributed by ${\bm{\Delta }}_2^*$ due to the augmentation of ${R_{{r_0}}}$, and the amplification of ${F_{2x}}$ is more pronounced in the smaller droplet. This phenomenon can be attributed to two reasons. First, the absolute values of ${F_{2x1}} = - {{\partial_x} \Delta _{2xx}^*}$ in both droplets increase with the growth of ${R_{{r_0}}}$, and the increased amplitudes are nearly identical [see Fig. \ref{F14}(b)] on account of the growth of the absolute values of ${{\partial ^2}{u_x}}/{\partial {x^2}}$. The second factor is related to the absolute value of  ${F_{2x2}} = - {{\partial_y} \Delta _{2xy}^*}$, which increases with the increase of ${R_{{r_0}}}$ within the larger droplet, but decreases with the increase of ${R_{{r_0}}}$ in smaller droplet [refer to Fig. \ref{F14}(c)]. Fig. \ref{F15} illustrates that when ${R_{{r_0}}} = 40/15$,  the area occupied by the vortex velocity in the small droplet is significantly smaller than when ${R_{{r_0}}} = 40/35$. In other words, the area occupied by the vortex velocity in the small droplet decreases as ${R_{{r_0}}}$ increases. This explains why the absolute value of ${F_{2x2}} =  - {{\partial_y} \Delta _{2xy}^*}$ decreases with the increase of  ${R_{{r_0}}}$ in the small droplet.
 \begin{figure}[htbp]
{\centering
\includegraphics[width=0.5\textwidth]{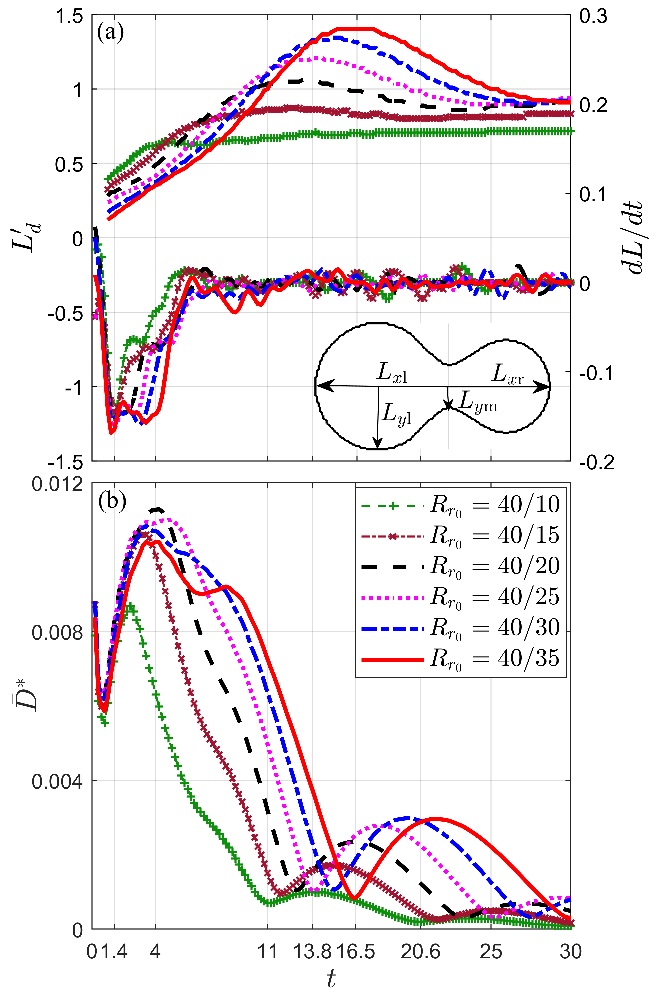}}
 \caption{\centering{Temporal profiles of (a) the ratio of minor and major axes $L'_d=(L_{y\rm m}+L_{y\rm l})/(L_{x\rm l}+L_{x\rm r})$, the slope of boundary length $dL/dt$, and (b) the total TNE strength $\bar D^*$ within the entire droplet  for $R_{r_0}=40/10, 40/15, 40/20, 40/25, 40/30, 40/35$.}}
  \label{F16}
\end{figure}

Figure \ref{F16}(a) displays the profiles of $L'_d=(L_{y\rm m}+L_{y\rm l})/(L_{x\rm l}+L_{x\rm r})$ (the left $y$ axis) and $dL/dt$ (the right $y$ axis) with different $R_{r_0}$. The physical meanings of $L_{y\rm m}$, $L_{y\rm l}$, $L_{x\rm l}$ and  $L_{x\rm r}$ are presented in the Fig. \ref{F16}(a). Fig. \ref{F16}(b) illustrates the time evolution curves of the total TNE strength $\bar D^*$ within the entire droplet for different $R_{r_0}$. From Fig. \ref{F16}(a), as a result of the increasing of the difference in curvature between large and small droplets and the decreasing mass inertia of small droplet with the increase of $R_{r_0}$, the larger $R_{r_0}$ is, the easier it is for small droplet to be absorbed by large droplet, the faster the coalescent process is, the shorter the oscillation period is, and the shorter the duration of the phase when the boundary length decreases the fastest is. Comparing Fig. \ref{F16}(a) with Fig. \ref{F16}(b),  it becomes evident that $\bar D^*$ still reaches its maximum at the stage when the boundary length changes most rapidly. Furthermore, after entering the damping oscillation stage, the valley points on the $\bar D^*$ evolution curve also correspond to the extreme points on the $L'_d$ curve. For $R_{r_0}>40/30$, the evolution curve of $\bar D^*$ becomes monotonically decreasing when $t_{D\rm max}<t<t_{l\rm m}$. This is because, in cases with $R_{r_0}>40/30$, the reduction in the amplitude of droplet oscillation and the diminished pushing effect of the left boundary on internal particles result in a monotonic decrease in the velocity gradient inside the droplet after reaching its maximum. More importantly, when the oscillation of droplet morphology is very weak, the evolution curve of $\bar D^*$ can also reflect the oscillation of the internal velocity of the droplet [see the green lines marked by `$+$' for $R_{r_0}=40/10$ in Fig. \ref{F16}].
\subsection{Entropy production during droplet coalescence}
Entropy increase is an important physical quantity in non-equilibrium thermodynamics. The heat transfer and flows of a thermal multiphase flow system can cause entropy increase in the system, which in turn can also affect and reflect the spatiotemporal evolution characteristics of the system. The entropy increase of a thermal multiphase flow system is contributed by entropy flow and entropy production. Entropy production plays a crucial role in non-equilibrium multiphase flow systems, and the spatiotemporal evolution characteristics of entropy production can reflect the process of system evolution\cite{LiRS}. For example, Zhang \emph{et al.}\cite{RN30} found that the entropy production rate increases with time at the spinodal decomposition stage and decreases with time in the domain growth stage, and the moment it reaches its maximum can be used as the second physical criterion for dividing these two stages. Chen \emph{et al.}\cite{Chen2022PRE} studied the evolution characteristics of entropy generation caused by two non-equilibrium effects (NOMFs and NOEFs) in Rayleigh-Taylor instability (RTI) systems and found that the first maximum value of the slope of the entropy product rate related to NOMFs can be utilized as an additional criterion for the evolution of RTI to enter the asymptotic velocity stage. Heat flux and dissipative stress are the two source terms that directly contribute to the entropy production of multiphase flow systems. The entropy production rate is given by\cite{LiRS}
\begin{equation}
\frac{{d{S_{{\rm{pr}}}}}}{{dt}} = \int {\left( {{{\bf{J}}} \cdot \bm{\nabla} \frac{1}{T} - \frac{1}{T}{\bm{\Pi}}:\bm{\nabla} {\bf{u}}} \right)} d{\bf{r}},
\end{equation}
where ${{\bf{J}}}$ and ${\bm{\Pi }}$  represent the heat flux and viscous stress tensor, respectively. According to  equation (6),  ${{\bf{J}}}= \bm{\Delta }_{3,1}^{\ast }+2\rho Tq \bm {\nabla} T$  and ${\bm{\Pi}}= \bm{\Delta }_{2}^{\ast }$, thus the expression of entropy production rate about  NOMFs and NOEFs is\cite{RN30}
\begin{equation}
\frac{{dS_{\rm{pr}}}}{{dt}} =  \int {\left[(\bm{\Delta }_{3,1}^{\ast }+2\rho Tq \bm{\nabla}T)\cdot{\bm {\nabla}  {\frac{1}{T}}}
- {\frac{1}{T}}{\bm{\Delta }_{2}^{\ast }}:\nabla {\bf{u}} \right]} d{\bf{r}}.
\end{equation}
These two entropy production rates are denoted as $ \dot S_{\rm NOEF}= \int {(\bm{\Delta }_{3,1}^{\ast }+2\rho Tq \bm{\nabla}T)\cdot{\bm {\nabla}  {\frac{1}{T}}}} d{\bf{r}}$ and $\dot S_{\rm NOMF} =  \int {- {\frac{1}{T}}{\bm{\Delta }_{2}^{\ast }}:\nabla {\bf{u}}} d{\bf{r}}$, respectively, i.e. $\dot S_{\rm pr}= \dot S_{\rm NOEF}+ \dot S_{\rm NOMF}$.

\begin{figure}[htbp]
{\centering
\includegraphics[width=0.5\textwidth]{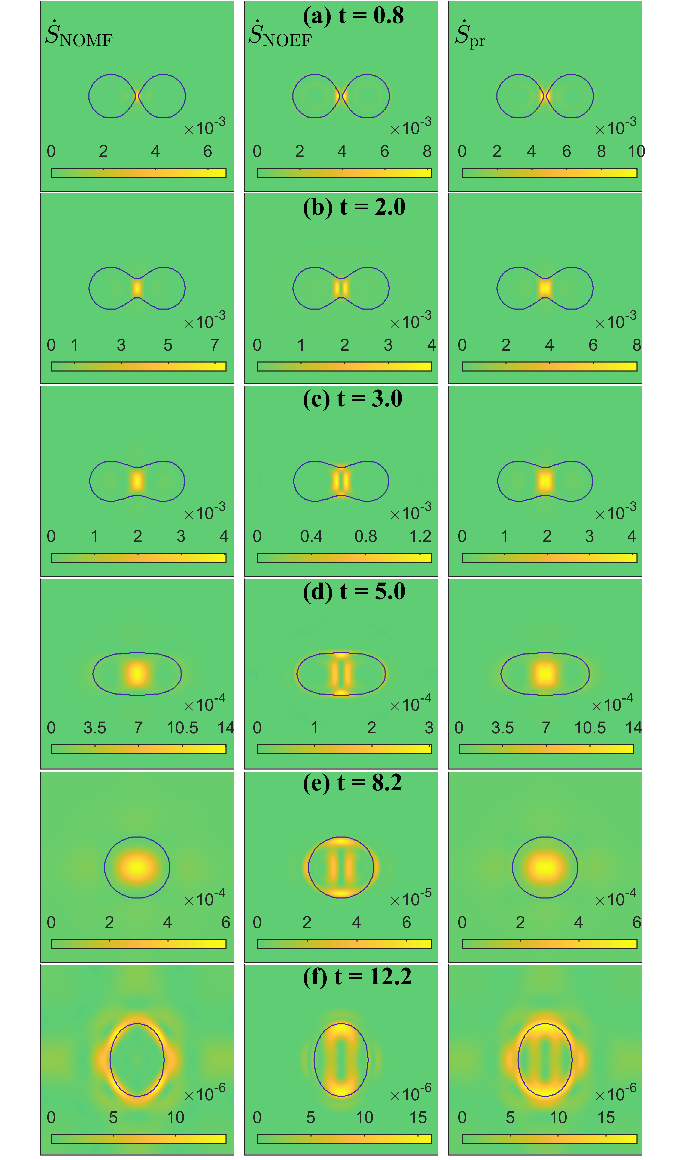}}
 \caption{\centering{Spatial distributions of $\dot S_{\rm NOMF }$ (first column),  $\dot S_{\rm NOEF }$ (second column), and  $\dot S_{\rm pr }$ (third column) at $t=0.8$, $2.0$, $3.0$, $5.0$, $8.2$,and $12.2$.  Here, $K = 0.0002$, $\tau  = 0.0025$,  $\Pr  = 0.5$, ${r_{0{\rm{l}}}} = {r_{0{\rm{r}}}} = 30$, and each submap encompasses the entire computational domain  ${N_x} \times {N_y} = 256 \times 256$.}}
 \label{F17}
\end{figure}

\begin{figure}[htbp]
{\centering
\includegraphics[width=0.45\textwidth]{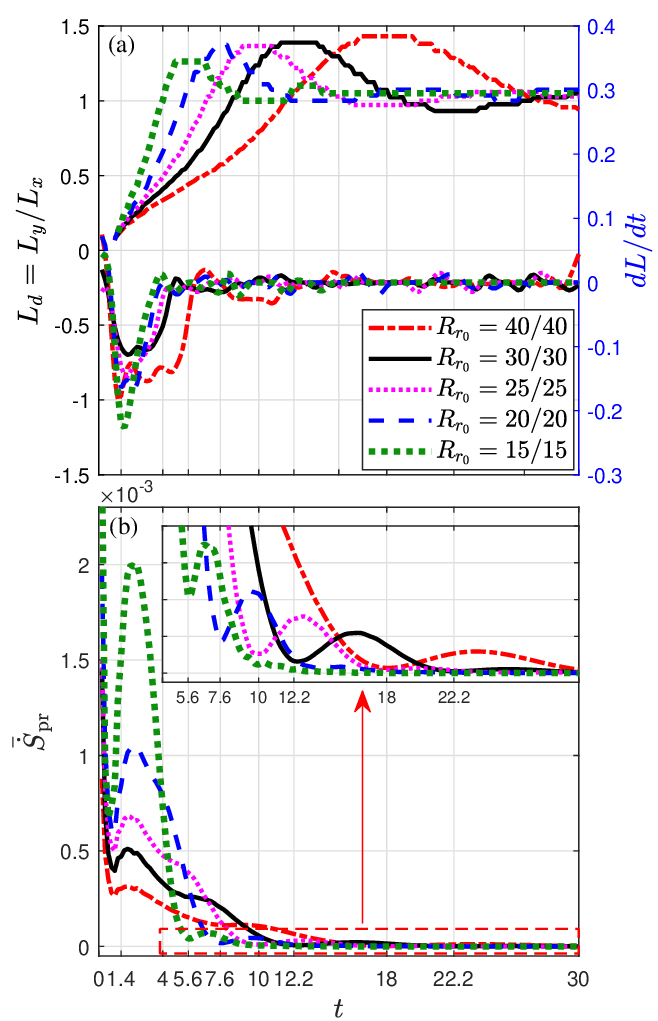}}
 \caption{\centering{Time evolutions of (a) the ratio of minor and major axes $L_d=L_{y}/L_{x}$ , the slope of boundary length $dL/dt$, and (b) the mean total entropy production rate within the whole droplet $\bar {\dot S}_{\rm pr}$ for $R_{r_0}=40/40, 30/30, 25/25, 20/20, 15/15$.}}
 \label{F18}
\end{figure}

The spatial patterns of $\dot S_{\rm NOMF }$, $\dot S_{\rm NOEF }$ and $\dot S_{\rm pr }$ at six characteristic instants are illustrated in Fig. \ref{F17}. Similar to TNE behaviors, the stronger entropy production rate initially appears at the liquid bridge between two liquid droplets. It is because the temperature gradient and velocity gradient in this area are greater, as seen in Fig. \ref{F1}(a), Fig. \ref{F1}(b) and Fig. \ref{F3}(a). More specifically, the entropy production rate contributed by the NOMFs, $\dot S_{\rm NOMF }$, which is dominated by the velocity gradient, is concentrated in the central region of the liquid bridge. On the other hand, the entropy production rate associated with by the heat flux, $\dot S_{\rm NOEF }$, spreads around the liquid bridge due to the temperature gradient mainly disperses in this area. Prior to the cut-through of the two droplets, peak of $\dot S_{\rm NOEF }$ is larger than that of $\dot S_{\rm NOMF }$, as depicted in Fig.  \ref{F17}(a). After the cut-through, the velocity gradient initially increases and then decreases with the expansion of the liquid bridge's radius. Consequently, the peak of $\dot S_{\rm NOMF}$ follows a similar evolutionary trend. However, the peak of $\dot S_{\rm NOEF}$ gradually decreases because molecules inside the droplet mix more evenly over time. Although the peak of the total entropy production rate, $\dot S_{\rm pr }$, also gradually decreases due to the greater decrease rate of  $\dot S_{\rm NOEF }$ compared to the increase rate $\dot S_{\rm NOMF}$, the spatial statistical average of $\dot S_{\rm pr }$, $\bar {\dot S}_{\rm pr}$, exhibits non-monotonic behavior [see Fig. \ref{F18}(b)]. This behavior is attributed to the expansion of the domain of the principal value of $\dot S_{\rm pr }$ (see the third column of Fig. \ref{F17}). It can be seen that the evolution trend of $\bar {\dot S}_{\rm pr }$ is mainly determined by $\dot S_{\rm NOMF}$, and that the spatiotemporal evolution of velocity gradient determines the evolution of $\dot S_{\rm NOMF}$. Here  $\bar {\dot S}_{\rm pr}= {{{\sum {\rho (x,y,t){\dot S}_{\rm pr }(x,y,t)}}}/{{\sum {\rho(x,y,t)}}}}$.
 \begin{figure}[htbp]
{\centering
\includegraphics[width=0.45\textwidth]{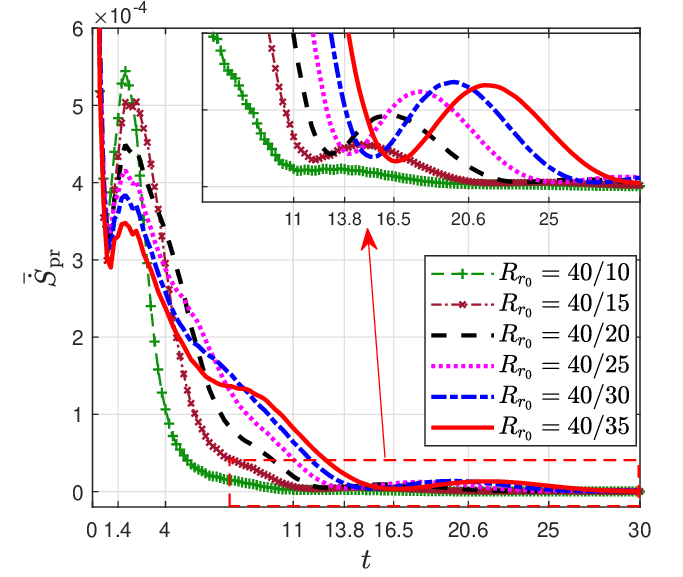}}
 \caption{\centering{Temporal profiles of the mean total entropy production rate within the whole droplet $\bar {\dot S}_{\rm pr}$ for $R_{r_0}=40/10, 40/15, 40/20, 40/25, 40/30, 40/35$.}}
 \label{F19}
\end{figure}
 \begin{figure*}[htbp]
{\centering
\includegraphics[width=0.8\textwidth]{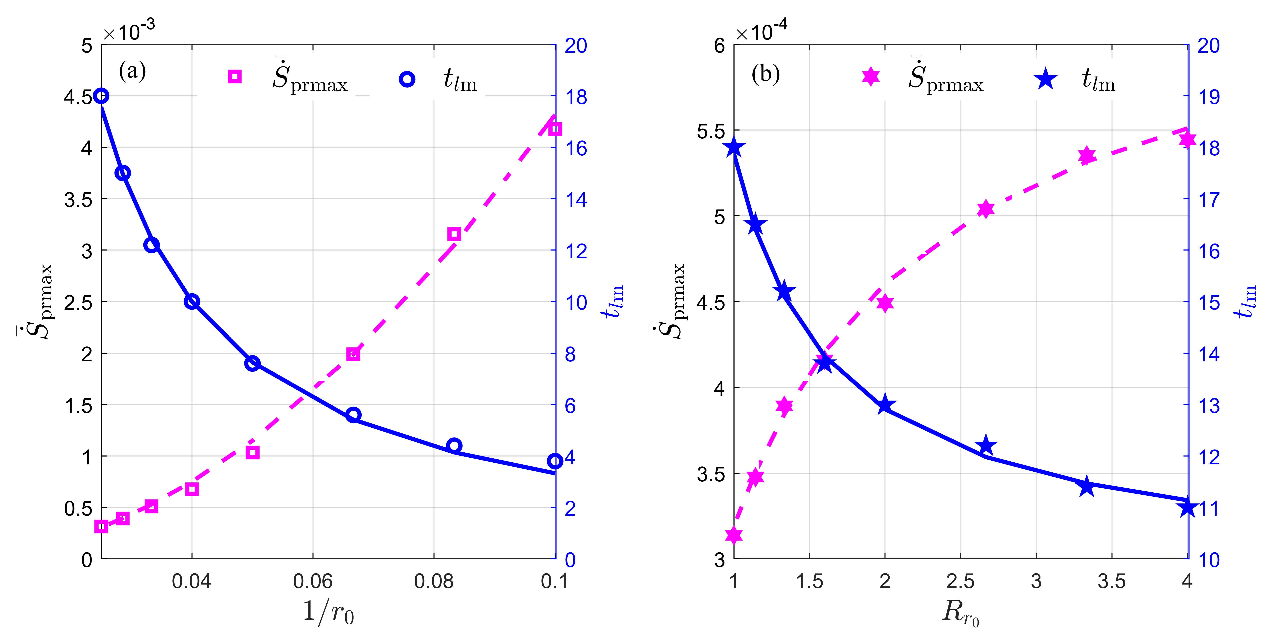}}
 \caption{\centering{(a) The effects of the initial radius on the maximum of the total entropy production rate within the whole droplet  $\bar {\dot S}_{\rm prmax}$ and the instant of $L_d$ being the maximum $t_{l\rm m}$; (b) The influences of the initial radius ratio
 ${R_{{r_0}}} = {r_{0 \rm l}} / {r_{0\rm r}}$
 on the $\bar {\dot S}_{\rm prmax}$ and the moment of the $L'_d=(L_{y\rm m}+L_{y\rm l})/(L_{x\rm l}+L_{x\rm r})$ reaching its maximum $t_{l\rm m}$.}}
 \label{F20}
\end{figure*}

Figure \ref{F18}(a) displays the profiles of $L_d$ (the left $y$ axis) and $dL/dt$ (the right $y$ axis) for $R_{r_0}=1$ with various initial radius ${r_0}$. Figure \ref{F18}(b) presents the time evolution curves of the total entropy production rate ($\bar {\dot S}_{\rm pr }$) within the entire droplet. From Fig. \ref{F18}(a), due to the increasing of the curvature droplet and the decreasing mass inertia of droplet with the decrease of the initial radius $r_0$, the lesser $r_0$ is, the faster the merging process is, the shorter the oscillation period is, and the shorter the duration of the fastest decreasing phase of the boundary length is. Interestingly,  from Fig. \ref{F18}(b) it is found that $\bar {\dot S}_{\rm pr }$ also reaches its maximum value at the stage when the changing rate of  liquid-vapor boundary length is the fastest, and that each instant of $\bar {\dot S}_{\rm pr }$ reaching the valley value corresponds to the moment when $L_d$ reaches its extremum, which is similar to the time evolution of $\bar D^*$. As shown in Fig. \ref{F19}, the aforementioned conclusion still holds for cases with $R_{r_0}>1$. The maximum value of $\bar {\dot S}_{\rm pr}$, denoted as $\bar {\dot S}_{\rm prmax}$, increases with the decrease of ${r_0}$ or the increase of $R_{r_0}$ [see the purple lines in Fig. \ref{F20}]. The functional relationships between them are expressed as $\bar {\dot S}_{\rm prmax} =0.35/r_0^{21/11}$ and $\bar {\dot S}_{\rm prmax}=(0.00066R_{r_0}-0.00021)/(R_{r_0}+0.41)$, respectively. The time for the system to reach $t_{lm}$ decreases with the decrease of ${r_0}$ or the increase of $R_{r_0}$ [see the blue lines in Fig. \ref{F20}], and the functional relationships among them are $t_{l\rm m} =0.21/r_0^{-6/5}$ and $t_{l\rm m} =(9.66R_{r_0}+1.94)/(R_{r_0}-0.35)$, respectively.
\section{Conclusions}\label{Conclusions}

In this paper, the two-dimensional discrete Boltzmann model for thermal multiphase flow is utilized to study the thermodynamic non-equilibrium (TNE) effects including entropy production rate during the coalescence of two initially motionless droplets.The relationships between the non-equilibrium effects,
the total entropy production rate and various physical
quantities associated with kinematics and dynamics are investigated
in detail.

During the thermal coalescence of two initially stationary droplets, in the coalescence direction, three primary forces govern the system evolution. Firstly, surface tension, resulting from the contribution of density and temperature gradient to the total pressure tensor, acts as the driving force for droplet coalescence.
Secondly, the force derived from the gradient of thermal pressure exhibits a dual role, hindering coalescence near the liquid bridge and promoting it farther away. Thirdly, the force contributed by non-organized momentum fluxes (NOMFs, $ {{\bm{\Delta }}_2^*}$), a typical non-equilibrium quantity, comprehensively restrains the coalescence process. However, the force contributed by $\Delta_{2xx}^*$ impedes the coalescence process, while the force contributed by $ \Delta _{2xy}^*$ accelerates it.

The spatial distributions of these TNE behaviors are mainly determined by the spatial distribution of the velocity field within the droplet.
When the radii of the left and right droplets differ, the asymmetry in the velocity distribution and velocity gradient within the droplets inevitably leads to an asymmetrical spatial distribution of non-equilibrium effects.
In the smaller droplet, the merging velocity, the total TNE strength $\bar D^*$ and the resistance contributed by ${\bm{\Delta }}_2^*$ in the direction of
coalescing are all greater, whereas the non-equilibrium components such as $ \Delta _{2xy}^*$ and forces related to shear velocity are smaller.
The larger the initial radius ratio, the larger the merging speed and the total TNE strength $\bar D^*$, and the shorter the moment of reaching the maximum value. A quadratic relationship exists between the ratio of the maximum merging velocity of the small and large droplets and the ratio of initial radius. The temporal evolution of the total TNE strength $ \bar D^*$ serves as a convenient and efficient physical criterion for distinguishing different stages of droplet coalescence: the total TNE strength $ \bar D^*$ inner the droplet reaches the maximum at the stage when the boundary length changes the fastest and each instant of the total TNE strength $ \bar D^*$ reaching the valley value is consistent with the moment that the droplet takes the longest elliptical shape. In the later stage of damping oscillation, even if the droplet's shape does not change anymore, the evolution curve of the total TNE strength $\bar D^*$ can still carefully reflect the velocity oscillation inside the droplet from mesoscopic level.

The entropy production rate is mainly contributed by the heat flux including the NOEFs and the NOMFs/viscous stress. The significant increase in entropy production rate initially takes place at the (quasi) contact point between two droplets, where strong temperature and velocity gradients exist. Prior to the merger of the two droplets, the primary cause of entropy production is the heat flow, whereas after the merger, the dominant factor in entropy production is the NOEFs/viscous stress.
Similar to the time evolution of the mean total TNE strength $ \bar D^*$, the spatial statistical average of entropy production rate ($\bar {\dot S}_{\rm pr}$) also reaches its maximum at the stage when the liquid-vapor boundary length changes most rapidly. Each moment when the entropy production rate $\bar {\dot S}_{\rm pr}$ reaches its minimum value corresponds to the time when the droplet assumes its longest elliptical shape. The smaller the initial radius or the larger the initial radius ratio is, the faster the droplet fusion process occurs, and the greater the maximum value of entropy generation rate $\bar {\dot S}_{\rm pr}$ is.

The fundamental investigations into various TNE behaviors and their influences, including the entropy production rate during the droplet coalescence process presented in this paper, enrich our comprehension of the mechanisms governing droplet coalescence from a mesoscale perspective. These findings provide novel and potential insights for accurately controlling droplet coalescence behavior in various engineering applications.

\begin{acknowledgments}
We acknowledge support from the National Natural Science Foundation of China
(Grant Nos. 11875001 and 12172061), Hebei Natural Science Foundation (Grant Nos. A2021409001 and A2023409003), Central Guidance on Local Science and Technology Development Fund of Hebei Province (Grant No. 226Z7601G), ``Three, Three and Three'' Talent Project of Hebei Province (Grant No. A202105005), the opening project of State Key Laboratory of Explosion Science and Technology (Beijing Institute of Technology) (Grant No. KFJJ23-02M),
Foundation of National Key Laboratory of Shock Wave and Detonation Physics (No. JCKYS2023212003),
Science Foundation of NCIAE (Grant Nos. KY202003 and GFCXJJ-2023-01),  Sponsorship Program for Introducing Overseas Scientists of Hebei Province (No. C20230121).

\end{acknowledgments}

\section*{Appendix}
\subsection{The discretization scheme for particle velocity space}
The specific discrete schemes of particle velocity, spacial and temporal derivatives couple together to make effect. Therefore, the specific discrete scheme of the particle velocity also needs to be reasonably selected according to the specific situation.
 The D2V33  model\cite{RN233} used in this work reads
\begin{equation}
{\mathbf{v}_{0}}=0,\mathrm{\ }{\mathbf{v}_{ki}}={v_{k}}\left[ {\cos ({\frac{i%
}{4}}\pi),\mathrm{\ }\sin ({\frac{i}{4}}\pi )}\right]{\bf e}_i ,
\end{equation}%
where $k=1,2,3,4$ indicates the $k$-th group of particle velocities whose
speed is ${v_{k}}$, $i=1,\cdots ,8$ is the direction of ${v_{k}}$ , and ${\bf e}_i$ is the unit vector in the $i$-direction.
It is important to emphasize that the choice of a DVM is determined based on the model's accuracy and stability in modeling applications.
The discrete equilibrium distribution function $f_{ki}^{eq}$ is
\begin{eqnarray}
\begin{aligned}
   f_{ki}^{eq} = \rho {F_k}\left[ {\left( {1 - {{{u^2}} \over {2T}} + {{{u^4}} \over {8{T^2}}}} \right)} \right. + {{{{\bf{v}}_{ki}} \cdot {\bf{u}}} \over T}\left( {1 - {{{u^2}} \over {2T}}} \right)  \cr \quad \quad
   \left. { + {{{{({{\bf{v}}_{ki}} \cdot {\bf{u}})}^2}} \over {2{T^2}}}\left( {1 - {{{u^2}} \over {2T}}} \right) + {{{{({{\bf{v}}_{ki}} \cdot {\bf{u}})}^3}} \over {6{T^3}}} + {{{{({{\bf{v}}_{ki}} \cdot {\bf{u}})}^4}} \over {24{T^4}}}} \right],
\end{aligned}
\end{eqnarray}%
with the global coefficients
\begin{eqnarray}
F_{k} &=&\frac{1}{%
v_{k}^{2}(v_{k}^{2}-v_{k+1}^{2})(v_{k}^{2}-v_{k+2}^{2})(v_{k}^{2}-v_{k+3}^{2})%
}\notag \\
&&[48T ^{4}-6(v_{k+1}^{2}+v_{k+2}^{2}+v_{k+3}^{2})T ^{3}+  \notag \\
&&(v_{k+1}^{2}v_{k+2}^{2}+v_{k+2}^{2}v_{k+3}^{2}+v_{k+3}^{2}v_{k+1}^{2})%
T ^{2}\notag \\
&&-\frac{v_{k+1}^{2}v_{k+2}^{2}v_{k+3}^{2}}{4}T ],
\label{fk_eq1}
\end{eqnarray}%
\begin{equation}
F_{0}=1-8(F_{1}+F_{2}+F_{3}+F_{4})\text{,}  \label{f0_eq}
\end{equation}%
where
\begin{equation}
\left\{ k+l\right\} =\left\{
\begin{array}{ll}
k+l & \text{ if }k+l\leq 4 \\
k+l-4 & \text{ if }k+l>4%
\end{array}%
\right. \text{.}  \label{kl_eq}
\end{equation}%

\subsection{The discretization scheme for spatial derivative}
The Fourier transform of function $f (x_j) $is defined as
\begin{equation}
\tilde{f}\left( k \right)=\Delta x\text{ }\sum\limits_{j=0}^{N-1}{f({{x}_{j}})}{{e}^{-{\bf{i}}kx}}.
\end{equation}
The inverse transformation is
\begin{equation}
f\left( {{x}_{j}} \right)=\frac{1}{L}\text{ }\sum\limits_{n=-{N}/{2}\;}^{{N}/{2}\;-1}{\tilde{f}\left( k \right){{e}^{\mathbf{i}kx}}},
\end{equation}
here, $k$ is the module of the wave vector $ \bf k $, $ \bf i $ is an imaginary unit, and $\tilde{f}(k)$ is the Fourier transform of the function $f(x)$. In equation (28), $k=2 \pi n/L $, $L=N \Delta x $ is the total length of the system and is evenly divided into $N$ parts. When $N$ is infinite or $\Delta x $ is infinitesimal, the left and right sides of the above two equations are completely equal. The derivative theorem of Fourier transform is
\begin{equation}
{{{\tilde{f}}}^{'}}\left( k \right)=\mathbf{i}k\times \tilde{f}\left( k \right),
\end{equation}
where ${{{\tilde{f}}}}\left( k \right)$ is the Fourier transform of $ \tilde {f}(x)$. This theorem provides an effective way to solve spatial derivative: firstly, transform $f (x) $ from the real space to complex space $\tilde{f}\left( k \right)$; secondly, multiply $\tilde{f}\left( k \right)$ with ${{\bf i}k}$; thirdly, perform an inverse transform on ${{{\tilde{f}}}^{'}}(k)$ to obtain the derivative ${{f}^{'}}(x)$.
Similarly,
the $n$th order derivative $f^{(n)}(x_{j})$ ($n\geq 2$), can be obtained through multiply $\tilde{f}(k)$ with $(ik)^{n}$,
$\tilde{f}^{(n)}(k)=(\mathbf{i}k)^{n}\times \tilde{f}(k)$.
To avoid the Gibbs phenomenon,
we first expand $K$ in Taylor series,
\begin{eqnarray}
k&=&\frac{\arcsin\left[ \sin\left( {k\Delta x}/{2}\; \right) \right]}{{\Delta x}/{2}\;}\notag \\
 &=&\frac{1}{{\Delta x}/{2}\;}\sum\limits_{n=0}^{\infty }{\frac{\Gamma (\frac{n}{2}){{\delta }_{0}}\Theta (n)\varepsilon (-1+n)}{\sqrt{\pi }n\Gamma (\frac{n+1}{2})}}{{\sin }^{n}}\left( {k\Delta x}/{2}\; \right).
\end{eqnarray}
where $\Gamma (n)=\int_{0}^{\infty }t^{n-1}e^{-t}dt$ is the Gamma function, $%
\Theta (n)= Mod[-1+n,2]$ is the Mod function and $\varepsilon (-1+n)$ is the
unit step function. Then, $k$ takes the form of an
appropriately truncated Taylor series expansion of $\sin(k\Delta x/2)$
to filter out more high frequency waves.

\subsection{The discretization scheme for temporal derivative}
The second-order Runge-Kutta scheme is employed to discretize the temporal derivative in the discrete Boltzmann equation
\begin{equation}
f_{ki}^{n+{1}/{2}\;}=f_{ki}^{n}-\left[ {{\mathbf{v}}_{ki}}\cdot \bm{\nabla}f_{ki}^{n}+\frac{1}{\tau }\left( f_{ki}^{n}-f_{ki}^{eq,n} \right)-I_{ki}^{n} \right]\frac{\Delta t}{2},
\end{equation}
\begin{equation}
f_{ki}^{n+1}=f_{ki}^{n}-\left[ {{\mathbf{v}}_{ki}}\cdot \bm{\nabla} f_{ki}^{n+{1}/{2}\;}+\frac{1}{\tau }\left( f_{ki}^{n+{1}/{2}\;}-f_{ki}^{eq,n+{1}/{2}\;} \right)-I_{ki}^{n+{1}/{2}\;} \right]\Delta t.
\end{equation}
where the superscripts $n$ and $n + 1$ represent two consecutive time steps, $\Delta t$ is the time step size, $I_{ki}$ denotes the external force term, and ${\mathbf{v}}_{ki}\cdot \bm{\nabla} f_{ki}$ signifies the convection term.
\section*{Data Availability}
The data that support the findings of this study are available from the corresponding author upon reasonable request.

\section*{References}
\bibliography{BIT}

\end{document}